\documentclass[reprint, amsmath,amssymb, aps,]{revtex4-2}

\usepackage{graphicx}
\usepackage{dcolumn}
\usepackage{xcolor}
\usepackage{bm}
\usepackage{hyperref}
\usepackage{amsmath}
\usepackage{orcidlink}
\usepackage[english]{babel}

\newcommand{\IM}{\mathrm{Im}}
\begin{document}

\title{Fiber-based scanning levitated nanoparticle force sensor near a metalized membrane}

\author{Andrew M. Dana\,~\orcidlink{0009-0007-8370-8328}}
    \altaffiliation{These authors contributed equally}
    \affiliation{Department of Physics and Astronomy, Northwestern University, Evanston, Illinois 60208, USA}

\author{Samuel J. Borden\,~\orcidlink{0009-0003-2539-4333}}
    \altaffiliation{These authors contributed equally}
    \affiliation{Department of Physics and Astronomy, Northwestern University, Evanston, Illinois 60208, USA}

\author{Nia Burrell}
    \affiliation{Department of Physics and Astronomy, Northwestern University, Evanston, Illinois 60208, USA}

\author{Chethn Galla\,~\orcidlink{0000-0003-3865-1346}}
    \affiliation{Department of Physics and Astronomy, Northwestern University, Evanston, Illinois 60208, USA}

\author{Alexey Grinin\,~\orcidlink{0000-0002-2838-7257}}
    \affiliation{Department of Physics and Astronomy, Northwestern University, Evanston, Illinois 60208, USA}

\author{Mark Nguyen\,~\orcidlink{0000-0002-7023-8642}}
    \affiliation{Department of Physics and Astronomy, Northwestern University, Evanston, Illinois 60208, USA}

\author{Kristina Menon}
    \affiliation{Department of Physics and Astronomy, Northwestern University, Evanston, Illinois 60208, USA}

\author{Shafaq Elahi\,~\orcidlink{0000-0002-3671-692X}}
    \affiliation{Department of Physics and Astronomy, Northwestern University, Evanston, Illinois 60208, USA}

\author{Katarina Boskovic Guy\,~\orcidlink{0009-0007-8534-9154}}
    \affiliation{Department of Physics and Astronomy, Northwestern University, Evanston, Illinois 60208, USA}

 \author{Andrew Laeuger\,~\orcidlink{0000-0002-8212-6496}}
    \affiliation{Department of Physics and Astronomy, Northwestern University, Evanston, Illinois 60208, USA}
 \author{Evan Weisman}
    \affiliation{Department of Physics and Astronomy, Northwestern University, Evanston, Illinois 60208, USA}
    
 \author{Gambhir Ranjit}
    \affiliation{Department of Physics and Astronomy, Northwestern University, Evanston, Illinois 60208, USA}

\author{Andrew A. Geraci\,~\orcidlink{0000-0001-7009-0118}}
    \email[Correspondence email address: ]{andrew.geraci@northwestern.edu}
    \affiliation{Department of Physics and Astronomy, Northwestern University, Evanston, Illinois 60208, USA}

\date{\today}

\begin{abstract}
    
We describe a fiber-based dual-beam trap for dielectric nanospheres with the ability to transfer the trapped particles into a retro-reflective standing-wave trap within micron-range distances of a metallic mirror surface. We illustrate its capability for three dimensional scanning force sensing over a several square micron area of an ultra-thin gold-coated silicon nitride mirror-membrane with a total thickness of approximately 370~nm. We also demonstrate three-dimensional laser feedback cooling and zeptonewton force resolution in high-vacuum at micron range from the membrane, showing promise for future improved tests of gravity at short distances and other surface force investigations. By using charged nanospheres, we expect the method may be useful for studying fluctuating patch potentials and electric field noise in the vicinity of a surface in the $\sim$ kHz to $\sim 100$ kHz frequency regime.
\end{abstract}

\maketitle

\section{Introduction} \label{sec:intro}

In recent years, levitated optomechanical systems have shown exceptional force sensitivities in free space \cite{ ranjit_zeptonewton_2016, hempston_force_2017, kawasaki_high_2020, ranjit_attonewton_2015,liang_yoctonewton_2023,monteiro_force_2020,novotnydrop,skrabulis2026nanomechanicalsensorresolvingimpulsive}, and there have been numerous advancements towards trapping and sensing forces near a variety of material surfaces \cite{grinin_optically_2026, montoya_scanning_2022, winstone_direct_2018, diehl_optical_2018, blakemore_three-dimensional_2019}. With such force or acceleration sensitivity, demonstrated ground state cooling \cite{tebbenjohanns_quantum_2021, delic_cooling_2020, piotrowski_simultaneous_2023, ranfagni_two-dimensional_2022}, and squeezing or quantum delocalization of the mechanical modes of such mesoscopic systems\cite{kamba_quantum_2025, kremer2026fastquantumsqueezingnanomechanical, rossi2024quantumdelocalizationlevitatednanoparticle}, levitated optomechanics has become 
a ripe testbed for fundamental physics \cite{Moore_2021}. Levitated optomechanical systems have been used or proposed for tests of gravitational forces \cite{geraci_short-range_2010, bose_spin_2017, marletto_gravitationally_2017}, observing nuclear decays \cite{mooredecay}, searching for dark matter \cite{tseng_search_2025,Hamaide2026}, searching for high frequency gravitational waves \cite{Arvanitaki:2013,aggarwal2022searching, winstone_optical_2022}, testing foundational aspects of the quantum to classical transition \cite{Bateman2014,ORI11_GM,Goldman2015} or quantum collapse models \cite{ORI11_GM,Vinante_2019}, and have been considered for applications in quantum information science, for example as local quantum memories in quantum networks \cite{Mancini2003, deplano2026stationaryentanglementlevitatedoscillator}. 

Among the many avenues of scientific inquiry using levitated systems, there are a variety of proposals to experimentally measure deviations to the gravitational inverse square law \cite{geraci_short-range_2010,blakemore_search_2021} or demonstrate gravitationally mediated entanglement which have garnered interest as a means of elucidating the fundamental nature of gravity \cite{bose_spin_2017, marletto_gravitationally_2017}. In many cases, these experiments require the levitated system to be in proximity to a nearby conducting surface as a means of shielding the sensor from electromagnetic forces which would overpower the underlying gravitational interaction of interest \cite{schut_micrometer-size_2024, schut_relaxation_2023}. However, this surface can itself produce unintended force noise due to effects such as patch potentials \cite{patch}, residual forces due to surface interactions with permanent dipole moments of the trapped particles \cite{blakemore_search_2021, rider_search_2016, rider_electrically_2019, blakemore_absolute_2020}, or extraneous light scattering \cite{blakemore_search_2021}. 

In this work, we show that a $300$~nm diameter SiO$_2$ sphere can be suspended at single micrometer range from a gold (Au) coated silicon-nitride (SiN) membrane in high vacuum using a fiber based retro-reflected optical standing wave trap. To achieve this, the sphere is first trapped in a counter-propagating dual-beam optical trap with two orthogonally polarized beams with foci sufficiently offset to provide a balanced radiation pressure force to levitate the sphere. The trapped sphere is then placed at micron range from a nanofabricated Au coated Si wafer which forms a retro-reflected standing wave. The wafer die has a knife-edge and a SiN membrane in the center. The sphere is originally trapped near the knife edge which first slices into the beam, and then the surface can be translated perpendicular to the laser beam axis to maneuver the sphere over the thin membrane region. This transition to the thin membrane must occur at a low enough pressure and laser intensity as to avoid thermal forces which can destabilize the trapped sphere. To maintain stable levitation of the sphere at $5\times10^{-6}$~mbar for multiple days, three dimensional optical linear feedback cooling of the center of mass motion is employed. 

To characterize the force noise due to surface effects such as fluctuating patch potentials and light scattering, we demonstrate three dimensional force sensing over an approximately $2.25~\mu$m  $\times ~2.25$ $\mu$m patch of the Au coated membrane with step sizes of $250$ nm at a moderate vacuum pressure of $0.67$~mbar. Finally, we show that this levitated optomechanical system is capable of achieving $10^{-21}$~N root-mean-squared (RMS) force resolution in high vacuum at $5\times10^{-6}$~mbar for a measurement taken over $10^5$~s, consistent with the expected thermal-noise-limited force sensitivity at this pressure. 

In this particle mass and distance regime, the system is sensitive enough to search into new parameter space of Yukawa corrections to the gravitational inverse square law and serves as a proof of principle of high vacuum trapping near conducting membrane shields for gravitationally induced entanglement experimental protocols.   We expect that this method may also be useful for examining Casimir forces as well as more generally for scanning force microscopy. For example, by using a nanosphere with a non-zero electric charge, the system can be operated as a scanning surface potentiometer, which could have utility for studying fluctuating patch potentials, and electric field noise in the ${\mathcal{O}}(1-10~\mu$m$)$ vicinity of a metallic surface in the frequency range of a few kHz to $\sim 100$ kHz.

\section{Experimental Setup and Methodology} \label{sec:setup}
\begin{figure*}
\centering{}
\includegraphics[width=0.95\textwidth]{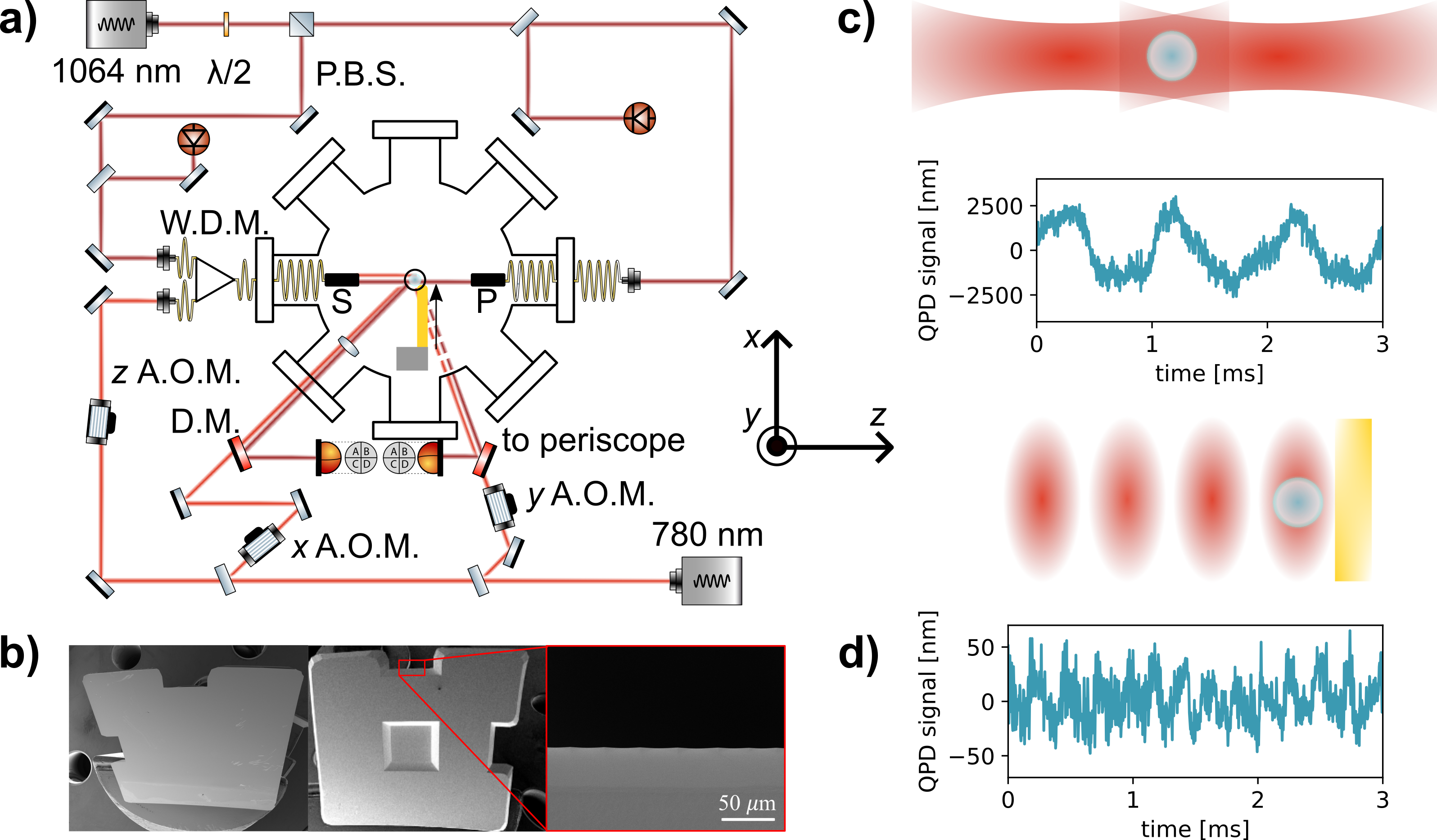} 
\caption{(a) A 1064~nm laser is used to initially trap  a 300~nm silica sphere in a fiber-based counter-propagating orthogonally-polarized dual-beam trap located in the vacuum chamber at the center of the diagram. A retro-reflected standing wave trap is then established by ``slicing" the dual-beam trap with a knife-edged gold-coated mirror, depicted by the gold rectangle in the center of the figure. Imaging lenses at the side and bottom (depicted by the open black circle in the center) of the chamber are used for detecting the sphere's motion. A $780$~nm laser is used for linear feedback cooling of the sphere's center-of-mass motion. (b) Scanning electron microscope images of the knife-edged gold-coated mirror, with the first panel showing the S-focus side, the second panel showing the P-focus side and the third panel showing a close-up of the knife-edge. (c) A 300~nm silica sphere starts in the counter-propagating orthogonally-polarized dual-beam optical dipole trap --- with the sphere's motion along the optical axis $z$ as imaged by the side QPD --- before (d) the 300~nm silica sphere is placed in the first antinode of the retro-reflected standing wave trap established by the gold-coated thin membrane. The sphere's motion along the optical axis $z$ as detected by the side QPD is also shown; note that the frequency of motion in the standing wave trap is much larger than in the dual-beam trap due to increased trap stiffness.}
\label{fig:experimentalsetup}
\end{figure*}

In this section, we describe the experimental setup and procedure for trapping a $300$~nm fused silica sphere a known distance away from a retro-reflecting mirror in order to perform force sensing experiments near a Au coated membrane. The experimental setup is shown in Figure \ref{fig:experimentalsetup}a. We initially trap a $300$~nm silica sphere in a fiber-based counter-propagating orthogonally-polarized dual-beam optical trap with offset foci so that the sphere’s position is precisely known. This dual-beam trap allows for larger mass spheres to be trapped than in a typical tweezer geometry. By then inserting a knife-edged mirror into the dual-beam trap — a process we will hereafter refer to as “slicing” — the sphere is transitioned into a retro-reflected standing wave trap at a repeatable distance away from the surface of the mirror. We discuss these steps in further detail.

\subsection{Fiber-based dual beam trap}

A free-space 1064~nm laser is sent through a $\lambda/2$ wave-plate and then split by a polarizing beam splitter (PBS) cube to create two equal-power orthogonally polarized (S \& P) beams. These two beams are then coupled into polarization maintaining fibers and feed-thru into a vacuum chamber where they couple to two diametrically opposed fiber-based pigtailed collimator packages. These two collimators each have an achromatic aspherical focusing lens (f=20~mm) and are rotated so that the S- and P-beams are orthogonally polarized relative to each other. The S-beam is polarized $\sim$45 degrees relative to vertical in the lab frame. This establishes a counter-propagating orthogonally-polarized dual-beam trap. The focuser carrying the P-polarized beam is clamped in position with the help of a V-block and a set screw, fixing its position. The focuser for the S-polarized beam is connected to a vacuum compatible three axis kinematic mount. The foci of the two counter-propagating beams — that have waists of $w_0 = 6.5~$$\mu$m — are offset by 80~$\mu$m so that even if the S- and P-branches have a slight power imbalance one robust trapping site is established, as discussed in Refs. \cite{ranjit_attonewton_2015,ranjit_zeptonewton_2016}. The $\lambda/2$ wave-plate is set so that both the S- and P-beams have an equal power, measured to be 160~mW, in the chamber. This offset-foci dual-beam trap is depicted in Fig. \ref{fig:experimentalsetup}c, along with a calibrated time-series trace of a $300$~nm sphere's motion in the optical axis direction ($z$).  A pulled-glass capillary micropipette attached to a piezoelectric transducer --- described in detail in Ref. \cite{grinin_localized_2026} --- is used to launch 300~nm silica spheres into the dual-beam trap. The fiber-based collimator packages are clamped to a stack of three vacuum compatible Newport Agilis piezo-motor driven 1D linear stages, similar to \cite{Mestres2015, harvey_nanomechanical_2022}; this allows for \textit{in-situ} control over the sphere’s location inside the vacuum chamber. Using a fiber-based interferometer (see Sec. \ref{sec:fiber_interferometer}), the average step size along the optical axis ($z$) is measured to be 250 nm.

\subsection{Distance metrology \label{sec:distance}}

Prior to slicing the knife-edged mirror into the dual beam and establishing a standing wave trap, we are able to precisely position the mirror's surface a distance $z_0$ behind the silica sphere. Without a sphere in the trap and with both beams hitting the mirror membrane, we are able to determine the locations of the S- and P-foci while translating the mirror along the optical axis $z$. This allows us to precisely position the mirror's surface at the trapping location, as discussed in further detail in Appendix \ref{sec:fiber_interferometer}. By carefully balancing the power between the two counter-propagating dual-beams, see Appendix \ref{sec:power_balancing}, the silica sphere is trapped exactly midway between the two offset foci. Using a fiber-based interferometer, we then position the mirror's surface a measured distance $z_0$ behind the sphere. 

The leading uncertainty in the sphere’s distance from the mirror’s surface is from the RMS motion of the sphere in the dual-beam trap prior to slicing. Using the equipartition theorem and the trap frequency along the optical axis of the dual-beam of $1.1$~kHz, we estimate the uncertainty in the sphere’s location to be $1.8~\mu$m. When incorporating other sources of measurement error, as discussed in Appendix \ref{sec:fiber_interferometer}, the total uncertainty is given by $\delta_{z_0} = 1.9~\mu$m. When reporting the sphere’s distance $z_0$ we use asymmetric error bars when the lower bound of the sphere’s position would be unphysical. 

\subsection{Slicing using a knife-edge mirror\label{sec:slicing}}
Once the sphere has been placed at a desired location relative to the mirror's surface using the 3D stage holding the dual-beam trap, a knife-edged Au-coated Si mirror is translated towards the dual-beam trap using an independent Newport Agilis stage traveling orthogonally to the optical axis ($x$). A scanning electron microscope image of the knife-edge is shown in Fig. \ref{fig:experimentalsetup}b. The $10$~mm $\times~10$~mm mirror of $0.5$~mm thickness is custom-made by Norcada, with a $2~$mm$\times~2$~mm region at the center that is $370$~nm thick and a $2~$mm-long region at the edge that is etched along the crystal axis to form the knife-edge. Cutting the dual-beam with this knife-edged mirror allows for an adiabatic transition between the dual-beam trap and the fully-sliced retro-reflected standing wave trap: the sphere’s root-mean-square velocity in the trap is $\sim$10~mm/s, while the maximum speed of the Newport Agilis stage is 0.5~mm/s. This transition is reversible --- that is, the sphere can be un-sliced from the standing wave trap back into the dual-beam trap. In our labeling, the sphere is trapped in the S-beam once it is in the retro-reflected standing wave trap. The process of inserting a mirror behind an optically levitated sphere to establish a standing wave trap was investigated in Ref. \cite{montoya_scanning_2022} which used a mirror without a knife-edge to slice into an optical tweezer. We find that the knife-edge feature on the mirror enables more repeatable transitioning between the dual-beam and standing wave traps.

\subsection{Retro-reflected standing wave trap}
Once the mirror is in place, the incident S-polarized beam establishes a standing wave trap due to interference with its retro-reflection, with the sphere trapped at an antinode. For a weakly focused beam, the position of the first anti-node is a distance of $\lambda/4$ away from the mirror's surface \cite{montoya_scanning_2022}, which is a distance of $266$~nm for $1064$~nm light. The rest of the trapping sites are spaced by $\lambda/2$, or $532$~nm for our $1064$~nm trapping light. The trapping potential is steepest along the optical axis. Figure \ref{fig:experimentalsetup}c depicts the retro-reflected standing wave trap as well as the motion of a silica sphere along the optical axis in the trap. In the standing wave trap, the surface of the mirror can be translated along $x$ and $y$ while the sphere stays stationary in the lab frame; the boundary conditions guarantee that the standing wave keeps the sphere at the same distance $z_0$ from the mirror during this translation \cite{montoya_scanning_2022}. For the same reasons, we can move the S-beam's focus independently of the sphere's trapping location, allowing us to place the S-focus closer to the sphere's location to create a stiffer trap. This allows for \textit{in-situ} frequency tuning without changing the trap power. A 370~nm thick Au coated SiN membrane is located at the center of the mirror, and we discuss the process of placing the sphere over this thin membrane in Section \ref{sec:thermophoresis}. This thin membrane can be seen in the second panel of Fig. \ref{fig:experimentalsetup}b as viewed from the P-beam side; the S-beam side is completely flat, as seen in the first panel of Fig. \ref{fig:experimentalsetup}b, in order to establish the retro-reflected standing wave trap. 

The position of the sphere is measured by imaging the sphere on two quadrant photodetectors (QPDs). Aspheric lenses (NA=0.24) image the sphere from the side (at a 45 degree angle relative to the trapping axis in the $(x,z)$ plane) and from the bottom (along $y$). The side QPD is therefore predominately sensitive to motion in the $y$ direction, and the bottom QPD is most sensitive to motion in the $(x,z)$ plane; however, due to the polarization of the trapping light, all three motional degrees of freedom are visible in both QPDs. 

In addition to measuring the displacement of the sphere, the QPD signals are also used to generate a velocity signal used for linear optical feedback cooling of the sphere’s center of mass motion \cite{li_millikelvin_2011}. This active feedback cooling is necessary to prevent photophoretic forces \cite{ranjit_attonewton_2015} --- due to uneven heating of the sphere --- from ejecting the sphere out of the trap at intermediate pressures while pumping to high vacuum \cite{Kiesel2013, Millen2014, Price2015}. Active feedback cooling also prevents large amplitude oscillations in the sphere's motion at high vacuum that can cause nonlinearities in the otherwise harmonic motion. Three separate beams of 780~nm light are amplitude modulated by acousto-optical modulators (AOMs) with the velocity signal in order to induce a damping term in the sphere’s motion. One 780~nm beam co-propagates in fiber with the 1064~nm S-polarized trapping beam after being mixed using a wavelength division multiplexer (WDM), while the other two beams are entirely in free-space and pass through the side and bottom imaging lenses. The cooling beam along the vertical $y$ direction is clipped by the presence of the mirror; we compensate by sending more power to the cooling laser in this direction. The velocity signal used to modulate the AOMs is generated by using the PyRPL software's \cite{neuhaus_python_2024} IQ demodulator module on a RedPitaya field-programmable gate array board to Lorentzian band-pass filter the position signal near the trapping frequencies and then phase-shift the filtered signal with the correct cooling phase, which is determined experimentally \cite{feldman_trapping_2025}.

\subsection{Thermophoretic and photophoretic force near a Au coated membrane \label{sec:thermophoresis}}

 To place the sphere over the Au coated SiN membrane, the mirror's surface is translated adiabatically over an approximately $4$~mm distance along the $x$ direction. The P-beam, which hits the uneven side of the Au coated wafer, is used to ensure that the sphere is located over the thin membrane by monitoring the amount of retro-reflected light coupling back into the optical fiber. It has been observed in this apparatus that certain pressures and laser powers prohibit stable trapping near the thin metalized membrane which is not the case for the thicker Au coated wafer.  We attribute this to the fact that the thicker substrate can dissipate heat from the laser into the bulk much more efficiently than the thin membrane. The laser heated membrane dissipates more heat through the surrounding gas in the vacuum chamber. This causes a thermal gradient in the levitated nanosphere's local environment generating a thermophoretic force \cite{talbot_thermophoresis_1980}, which for certain values of pressure and laser power, can be strong enough to overcome the gradient forces confining the sphere. Notably, this is not an effect due to the internal temperature of the sphere which is heated by the trapping laser. Studies of the internal temperature of levitated spheres and the photophoretic force have been studied in detail in other works for silica \cite{chang_cavity_2010,millen_nanoscale_2014,ranjit_attonewton_2015}, nanodiamonds \cite{frangeskou_pure_2018, riviere_thermometry_2022} and rare-earth-doped nanospheres \cite{rahman_laser_2017,zhang_determining_2023}. We are particularly interested in studying the thermophoretic force around pressures in the $1$~mbar regime and greater: it is at this pressure where we initially place the sphere over the metalized membrane. 

To model the contributions from both photophoretic and thermophoretic forces as a function of pressure $P$, we use the interpolation formula \cite{fuchs_aerosols}
\begin{equation}\label{eq:Ft}
    F_T = -\frac{\pi r^2 \eta \sqrt{\frac{\alpha_g R_g }{M T} } \Gamma_i}{\frac{P}{P_0}+\frac{P_0}{P}}
\end{equation}
where $r$ is the sphere radius, $\eta$ is the gas viscosity, $\alpha_g$ is the accommodation coefficient, $R_g$ is the gas constant, $M$ is the molar mass, $T$ is the temperature of the environment, $P_0$ is the mean free path pressure and $\Gamma_i$ is the thermal gradient.  For the case of photophoresis, it is the thermal gradient across the sphere due to laser heating which matters, whereas for thermophoresis, it is the thermal gradient of the gas surrounding the sphere which matters. We note that while this is a good interpolation formula across the Knudsen regimes for modeling photophoretic forces, this formula is only approximate for thermophoretic forces in the continuum and transition regimes and does not hold well into the free molecular flow regime. In this regard, we are primarily interested in the qualitative behavior of the thermophoretic force at relevant pressures as a means of considering the plausibility that this effect accounts for the experimental observation.

To model both of these forces, we must find the thermal gradient $\Gamma_i$ for each case. For photophoresis, we look at the balance of power radiated and absorbed by the sphere due to black body photons, trapping laser photons and gas collisions. From this balanced equation, the internal temperature gradient across the sphere as a function of pressure can be estimated. For a detailed discussion of this, see Appendix \ref{sec:photophoresis}.

\begin{figure}[h!]
\includegraphics[width=\linewidth]{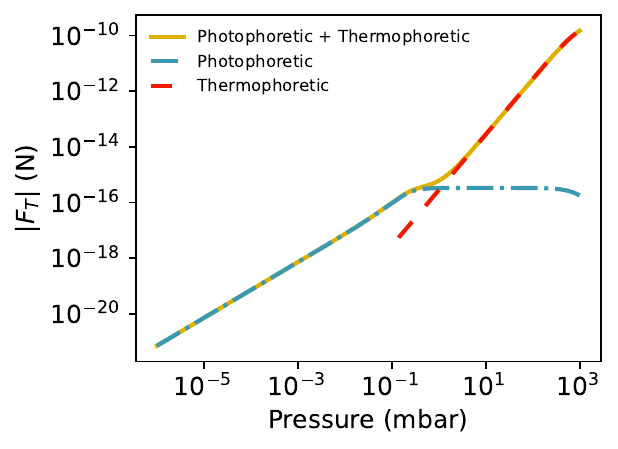}
\caption{Calculated thermophoretic and photophoretic force on a $300$~nm silica sphere in proximity to a laser heated metalized membrane. While the photophoretic force plateaus in the higher pressure regime and falls off at lower pressure, the thermophoretic force is significantly greater than the optical gradient force confining the sphere until pressures below $1$ mbar.}
\label{fig:phoreticforces}
\end{figure}

To approximate the thermal gradient across the gas in the local environment of the sphere, we take the temperature of the membrane heated by the laser $T_h$, the ambient temperature far from the membrane $T_c$, and estimate the gradient as $\frac{T_h - T_c}{n l_{mfp}}$ where $n$ is the average number of collisions for the gas to thermalize and $l_{mfp}$ is the mean free path as a function of pressure. Here $n$ is taken to be $5$ as the translational modes of air molecules thermalize very rapidly. We conduct finite element simulations using the COMSOL Multiphysics$^{\circledR}$ software \cite{noauthor_comsol_nodate} to quantify the temperature of the laser heated membrane $T_h$ at two different locations, the center and near one edge. At these two locations, the maximum surface temperatures are $T_{center} \approx420$~K and $T_{edge} \approx 370$~K. The value in the center is used in the subsequent calculations to serve as an upper bound on thermophoretic forces near the metalized membrane. 

With the thermal gradients for both cases, Eq. \ref{eq:Ft} can be used to calculate the magnitude of both photophoretic and thermophoretic forces on a nanosphere $\approx 1$~$\mu$m from the center of a metalized membrane as shown in Figure \ref{fig:phoreticforces}. Considering that the weakest gradient force holding the nanosphere in the optical trap, which is in the transverse directions, is approximately $10^{-14}$ N, this figure suggests that the sphere will experience significant thermophoretic forces until pressures below $\sim1$ mbar.  This qualitative model agrees with our experimental observations that spheres will stay trapped over the thicker Au coated surface across a large pressure and laser power regime, while to stably trap spheres over the metalized membrane, the typical laser power and pressure must be reduced. This demonstrates an aspect of trapping over metalized membranes which has not been seen in cases of trapping over dielectric membranes. In practice, this effect is circumvented by transitioning spheres to the metalized membrane at $0.67$ mbar or lower, with a laser power of $\approx 100$~mW, and operating near the edge of the membrane to reduce the thermal gradient in the surrounding gas. We note that smaller membranes could also be used to reduce this effect.

\section{Results and Discussion} \label{sec:results}
This section describes the results of numerous studies using $300$~nm fused silica spheres trapped at micron range from a Au coated SiN membrane or Au coated Si wafer in a retro-reflected standing wave trap. The net charge of multiple spheres was calibrated using a tungsten rod parallel to the grounded Au surface generating an electric field along the optical axis of the trap. Three dimensional scanning force sensing is demonstrated by scanning the trap site over a two dimensional grid along the membrane surface. The RMS force experienced by each translational degree of freedom of the sphere is measured over a surface area an order of magnitude greater than the sphere size. At a high vacuum, RMS force resolution of one mechanical degree of freedom is measured over $10^5$ seconds with a nanosphere at single micron range from the Au coated membrane. Analysis of this data was conducted to study the sources of force noise in the detection set up. All experiments were done with the sphere levitated in the S-beam. Interpretation and significance of the results are discussed in detail.

\subsection{Force and charge calibration}\label{sec:force_and_charge_calibration}

Once a $300$~nm silica sphere is trapped at a desired distance $z_0$ from the mirror and placed above the thin membrane, before applying feedback cooling and pumping to high vacuum, two separate datasets are acquired at $0.67$~mbar: one to calibrate the RMS force on the sphere and the other to determine its net electric charge. At this pressure, the sphere is in thermal equilibrium with the surrounding gas. We take $30$~s of force calibration data and use the equipartition theorem for each motional degree of freedom to convert the QPDs' voltage signals to displacements \cite{hauer_general_2013, hebestreit_calibration_2018, li_fundamental_2013}.

\begin{figure}
    \centering
    \includegraphics[width=\linewidth]{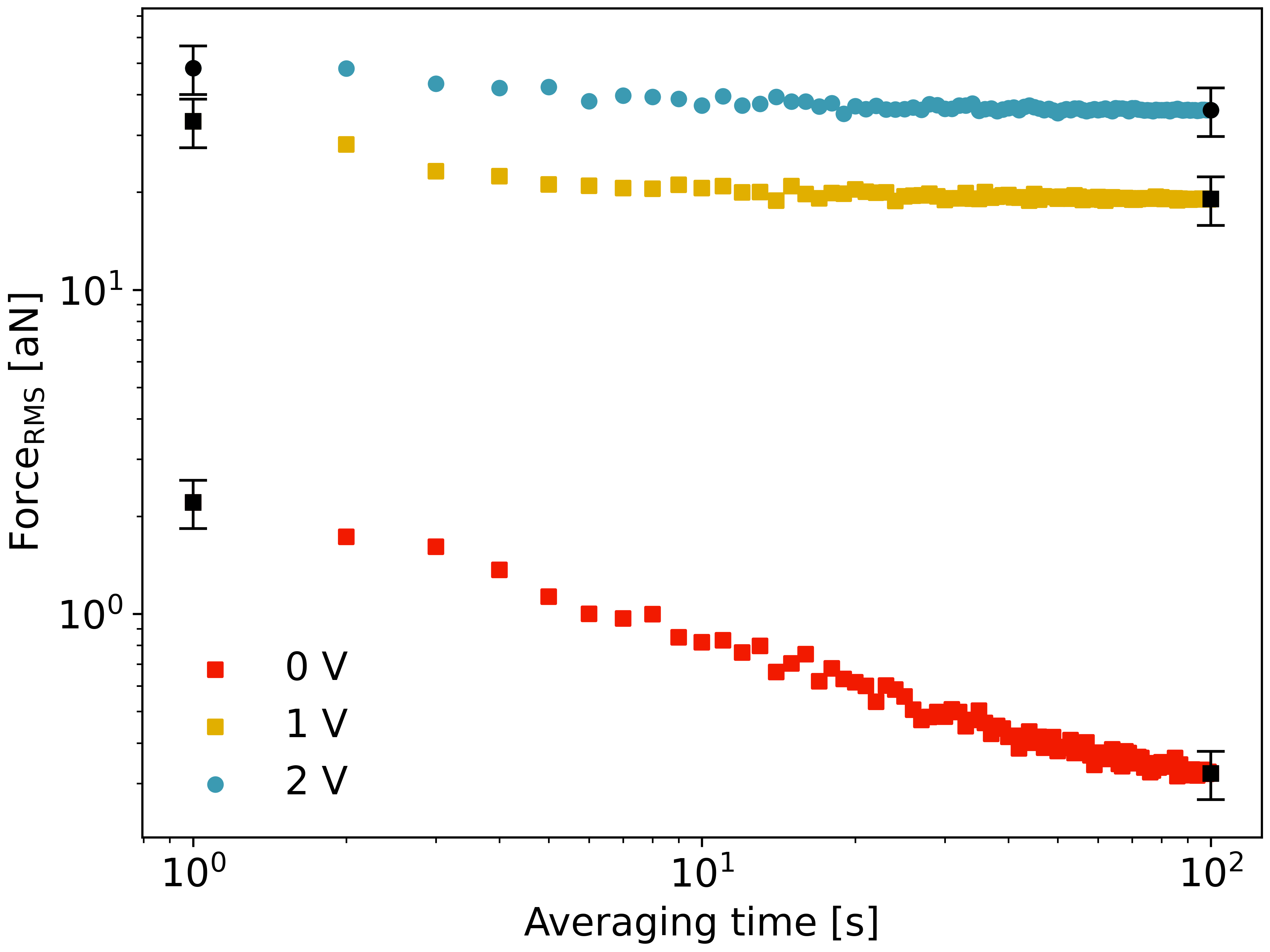}
    \caption{RMS force measured along the $z$-direction by applying on-resonance oscillating electric fields ($f=108$~kHz) of $28.5$ and $57.0$~V/m to a 300~nm silica sphere with a charge of $|q|=6e^-$ while trapped $1.6_{-1.3}^{+1.9}~\mu$m away from the $0.5$~mm thick part of the mirror's surface while at $9\times 10^{-6}$~mbar. Also shown is the force measured on the sphere in the absence of any driving force.}
    \label{fig:chargecalibration}
\end{figure}

The silica spheres can carry net electric charge --- due to triboelectric charging from passing through the glass capillary when launching \cite{siegel_optical_2025} --- which we can measure by applying a known oscillating electric field and measuring the resulting RMS force on the sphere. We apply an oscillatory electric field by sinusoidally driving voltage through a tungsten electrode placed parallel to the plane of the mirror, which is held at ground. The electric field is largest along the $z$ direction for this geometry; therefore, we drive the voltage on the electrode with a sine-wave on resonance with the sphere's motion in the $z$ direction. The electric field is modeled using the COMSOL Multiphysics$^{\circledR}$ software \cite{noauthor_comsol_nodate}: for $10$~V applied to the tungsten electrode, the electric field along the $z$ direction $1~\mu$m away from the mirror's surface is simulated to be $570$~V/m. We measure the RMS force at the applied voltage signal's frequency at a variety of applied field strengths to measure the net charge using $F=|q|E$ \cite{ranjit_attonewton_2015, ranjit_zeptonewton_2016}. We have measured that the net charge on the spheres ranges from $1-10$ elementary charges, and that $\sim20\%$ of the trapped spheres are uncharged. An example of measuring the net charge on a sphere by driving on-resonance is shown in Fig. \ref{fig:chargecalibration}. 

\subsection{Surface force microscopy over a metalized membrane} 
\label{sec:surface_scan}

\begin{figure*}
\centering{}
\includegraphics[width=\textwidth]{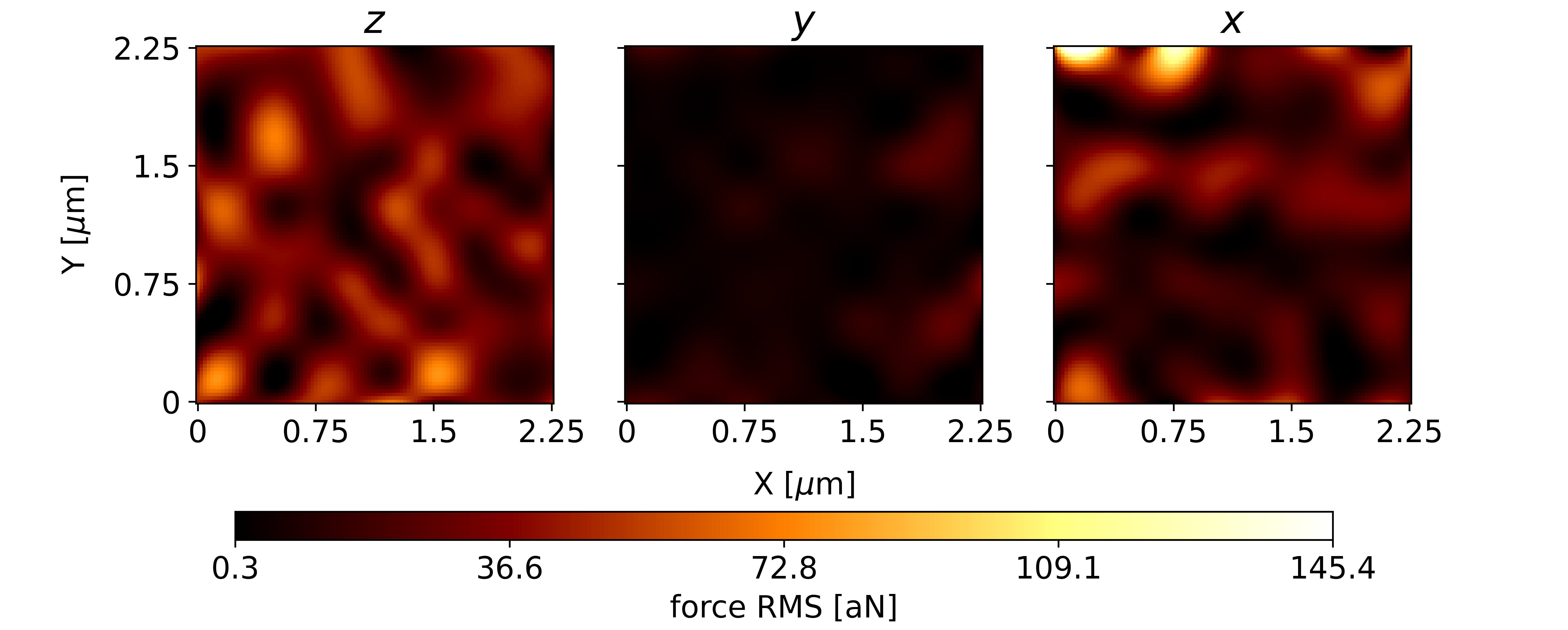} 
\caption{Root-mean-squared force measurements after 10 seconds of averaging for all three motional degrees of freedom measured with a 300~nm silica sphere $0.6_{-0.3}^{+1.9}~\mu$m away from the metalized membrane at a pressure of $0.67$~mbar. The sphere was scanned over a $10\times 10$ grid of points with each step corresponding to $250$~nm of stage motion. A 2D cubic spline interpolation between grid points was used for data visualization. All values below $72.8$~aN are consistent with thermal noise fluctuations. The RMS force could be on average larger in the axial $z$ direction due to stray electric fields terminating on the conducting surface and influencing the charged sphere's motion. For this particular scan, the $x$ direction showed a larger mean force after $10$~s of averaging than the $y$ direction, however we expect that on average these directions should exhibit similar levels of force noise.}
\label{fig:membranegrid}
\end{figure*}

The distance between the sphere and the surface is fixed when the retro-reflecting trap is formed. With a sphere placed $0.6^{+1.9}_{-0.3}$~$\mu$m from the metalized membrane, its location is scanned over an approximately $2.25 \times 2.25$~$\mu$m grid with step size of $250$~nm at a pressure of $0.67$~mbar. A range of surface to sphere distances and grid sizes have been tested;  more examples can be found in Appendix \ref{sec:additionalscans}. Figure \ref{fig:membranegrid} shows the measured RMS force over ten seconds at each point on this grid for the three translational degrees of freedom of the levitated sphere. Lorentzian fits to the thermal motion at each point are used to determine frequency and damping rate. The trap frequencies in this experiment were $(f_x, f_y, f_z) = (5522\pm6, 6049\pm2, 103411\pm7)$~Hz. We apply the same force calibration, taken at the origin of the scan, to every point in the grid: the alignment of the QPDs do not significantly change over the scan \cite{burrell_force_2025}. The relative force noise along each degree of freedom is calculated. The thermal force noise power spectral density is given by 
\begin{equation}\label{eq:SFF}
    S_{FF}^{th} = 4k_BT m\Gamma
\end{equation}
where $T$ is the effective bath temperature, $m$ is the sphere's mass and $\Gamma$ is the effective damping rate \cite{braginskyweakforces}. We considered the minimum detectable force for a given measurement bandwidth $b$ defined as
\begin{equation}
    F_{min} = \sqrt{S_{FF}^{th} b}.
\end{equation}

Ten seconds of data is taken at each grid point, and all values less than $72.8$~aN are consistent with thermal noise fluctuations within $3\sigma$ of the minimal detectable force. The mean force after 10~s of averaging for each degree of freedom is $(23.8\pm 5.6$, $6.7\pm 1.5$, $31.3\pm5.8)$ aN for $x, y$ and $z$ respectively, and the average minimal detectable force is $(13.5,10.5,24.5)$ aN. 
For the area of the surface studied here, the $x$ direction showed a larger mean force after $10$~s of averaging than the $y$ direction, indicative that in this case there were points on the surface where $x$ experienced significantly larger forces than $y$, however we expect that given more scans of the different parts of the surface, these directions should exhibit similar levels of force noise on average. We posit that the mean sensitivity being the largest in the $z$ direction could be due to the fact that in the grounded plate geometry, stray electric field lines will always terminate normal to the surface and will not significantly affect the transverse directions. The dominant source of uncertainty comes from a $5\%$ uncertainty on the sphere's radius \cite{ranjit_attonewton_2015, ranjit_zeptonewton_2016, harvey_nanomechanical_2022}, with subdominant contributions from the quality factor.  
To better understand the possible origins of measured values larger than $72.8$~aN, effects such as scattered light off the surface from local defects and contaminants are considered by measuring the correlation of the force noise with the backwards reflected laser power. At each grid point, the back reflected trap power is measured on a photodetector outside of the vacuum chamber as depicted in Figure \ref{fig:experimentalsetup}a. We assume that if a light scattering defect or contaminant on the surface induces excess force noise, it would correlate with a drop in voltage on the photodetector. No correlation between the photodetector voltage and force noise on any of the mechanical degrees of freedom is measured. More details can be found in Appendix \ref{sec:backreflectedpower}. It is clear that depending on the location of the sphere over the surface, the force sensitivity can vary for different degrees of freedom, and at some points the variations are significant enough to be considered measured forces as opposed to force noise due to the thermal background. We conclude that at these points, the measured forces are possibly due to interactions with the nearby membrane which are not due to localized surface defects or contaminants and rather consequences of other surface interactions such as fluctuating patch potentials. In many cases, such as short range tests of gravity, it is desired to use the spot over the surface and degree of freedom with the least amount of surface force noise. Alternatively, as a scanning force sensor, the high sensitivity of the levitated sphere can also be used as a probe of localized patches on the surface. As a rough comparison, state of the art nanomechanical cantilevers have demonstrated force sensitivities of $\sim1$~aN/$\sqrt{\mathrm{Hz}}$ at ultra-high vacuum pressures and cryogenic temperatures \cite{stowe_attonewton_1997, mamin_sub-attonewton_2001, tao_single-crystal_2014}. Furthermore, conventional atomic force microscopes (AFMs), relying on cantilever beams, typically have worse sensitivity, but routinely achieve sub-nm lateral and sub-Angstrom axial spatial resolution while operating at surface-tip distances of ~$0-10$~nm from the surface substrate.

Taking SEFN as an example, the lateral and axial spatial resolutions of this apparatus are estimated by considering the induced surface charge density and the thermal noise limited displacement sensitivity on resonance. For values of axial force sensitivity shown in Figure \ref{fig:chargecalibration} this optically levitated force microscope could potentially achieve approximately $30$~pm displacement resolution on resonance. Better axial resolution could be achieved by, for example, operating at lower pressures. Lateral resolution is limited by the size of the probe and the smallest achievable sphere to surface distance constrained by the laser wavelength. For more details on these estimates, refer to Section \ref{sec:resolution}. We note that the three dimensional vector nature of this force sensor makes this distinct from the operational mode of a conventional AFM which typically measures only the force perpendicular to the surface. Our method can measure the component of force in the plane of the surface due to a nearby feature. 

For experiments aiming to mitigate surface induced noise to obtain the best sensitivity to other forces, such as gravity, Figure \ref{fig:membranegrid} shows that the transverse degrees of freedom exhibit the least amount of noise on average. Additionally, neutral spheres are less directly susceptible to electric field noise. Towards this goal, the following section studies the RMS force resolution of a sphere at single micron range from the membrane in high vacuum for longer averaging times at a single location. This study looks at just the vertical degree of freedom in order to achieve the best force sensitivity and shows that thermal noise limited force sensitivity is achievable on resonance for proper choice of sphere location over the membrane.

\subsection{Force noise over the metalized membrane in high vacuum}\label{sec:HV_membrane_results}
\begin{figure}[h]
\includegraphics[width=\linewidth]{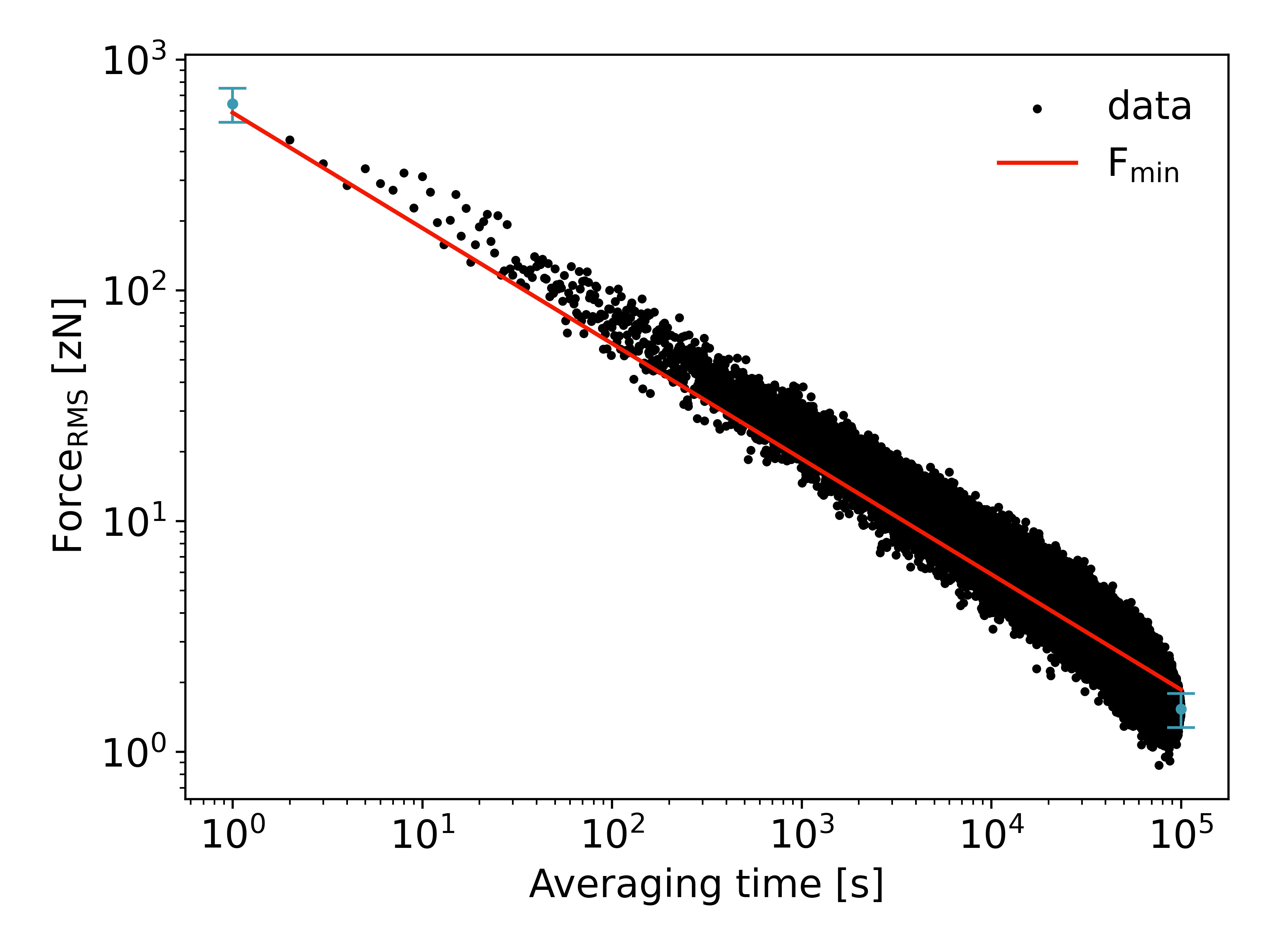}
\caption{Measured on-resonance RMS force resolution (in black) and the calculated theoretical minimal detectable force (in red) in the vertical ($y-$) direction for a 300~nm silica sphere $1.0_{-0.7}^{+1.9}~\mu$m away from a gold-coated metalized membrane. Data were collected for $1\times10^5$ seconds at a pressure of $5\times 10^{-6}$~mbar. Zeptonewton levels of force resolution are achieved.}
\label{fig:frms}
\end{figure}

To reduce force noise on the levitated sphere due to gas collisions (thermal noise), it is necessary to reduce the pressure in the experimental chamber. Furthermore, longer measurement times will improve the minimum detectable force with a $b^{1/2}$ scaling. As such, we analyzed the RMS force resolution of just one degree of freedom for a sphere located $1.0^{+1.9}_{-0.7}$~$\mu$m from the Au coated membrane for $10^5$~s. The chamber was held at a pressure of $5\times 10^{-6}$~mbar for the duration of this experiment.  Experimental observations have shown that spheres can stay trapped near the membrane in high vacuum for multiple days at a time without the need to adjust the feedback cooling parameters. The $10^5$~s of data shown here account for a segment of the data set in which the frequency of oscillation remains within a linewidth of the Lorentzian peak centered at $f_0=4536.5\pm2.5$~Hz. For each second of data, the peak amplitude is used to estimate the force as a function of the measurement time. Figure \ref{fig:frms} shows that over this $10^5$~s of averaging time, a force resolution of $F_{min} = (1.5\pm0.3)\times 10^{-21}$~N with a force sensitivity of $\sqrt{S^{th}_{FF}} = (642.8\pm107.8)\times 10^{-21}$~N/$\sqrt{\mathrm{Hz}}$ is achieved, and the data scale reasonably well with the theoretical expectation for a sphere with CoM temperature of $T_{\mathrm{CoM}} = 370\pm 80$~mK and damping rate $\Gamma = 95.7\pm2.5$~Hz. The theoretical thermal noise limited force sensitivity with these measured values is $588.3\pm78.9$~zN$/\sqrt{\mathrm{Hz}}$. Furthermore, we do not observe a plateau of the RMS force over this time scale indicating that there is no significant technical source of noise preventing an improved force resolution by averaging for longer times. This result is demonstrative evidence that the high force sensitivity of levitated nanospheres can be retained even when placed at micron range from a conductive membrane, as such geometries are often limited by noise due to interactions with the material surface (e.g. fluctuating patch potentials and other surface contamination). 

These experimental results open pathways towards future experiments concerned with forces between a dielectric sphere and an additional mass on the other side of the conducting membrane. We have shown that noise induced from the surface can be sufficiently mitigated in order to use it as an effective electromagnetic shield between the masses on each side. Importantly, the single zeptonewton force resolution achieved near the membrane, without plateauing over the measurement bandwidth, indicates that the nearby conducting surface does not significantly limit the force sensing capability at a level relevant for studying new parameter spaces of fundamental physics. Future experiments could include, but are not limited to, studies of the Casimir-Polder force, interactions of a levitated nanosphere to a nearby surface or to a secondary mass on the other side of the membrane \cite{schut_micrometer-size_2024, bose_spin_2017, jakubec_decoherence_2025}, surface electric field noise (SEFN) studies in the $1-100$~kHz frequency and $1-10$~$\mu$m distance regimes and the gravitational inverse square law in the single micron regime\cite{geraci_short-range_2010, blakemore_search_2021}. The following section analyzes the strength of various noise sources and offers strategies towards improving force sensitivity in the future.

\subsection{Noise analysis}

To gain insight into the primary sources of noise limiting this apparatus, we analyze the displacement and force noise for one second of data taken from the $10^5$~s studied in Section \ref{sec:HV_membrane_results}.  We primarily focus this analysis to the leading sources of noise, thermal noise due to gas collisions and detector noise. We also consider backaction noise, which is the next leading fundamental source of noise, and place an upper limit on potential SEFN. In the presence of a stationary fluctuating force, the thermal force power spectral density is given by Eq. \ref{eq:SFF}.  This is related to the displacement power spectral density $S_{qq}$ via the mechanical susceptibility $\chi(\omega)= \frac{1}{m (\omega_0^2 - \omega^2 + i\Gamma\omega)}$ such that
\begin{equation}
    S_{qq}(\omega) = |\chi(\omega)|^2S_{FF}^{th}
\end{equation}
\begin{figure}[h!]
\includegraphics[width=\linewidth]{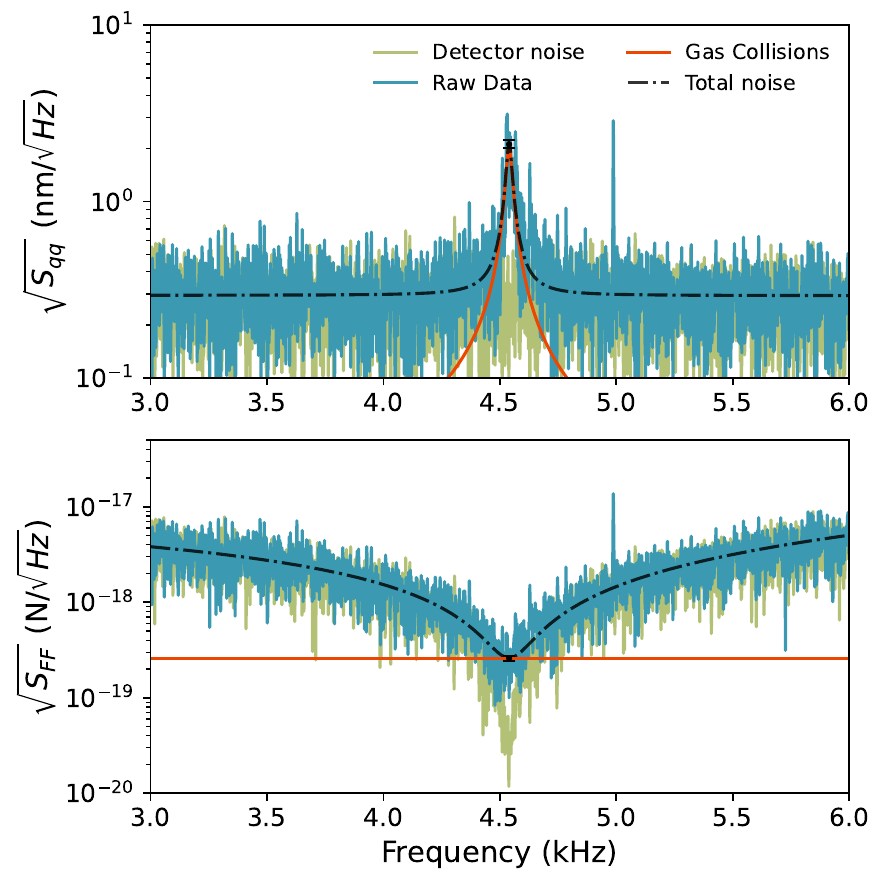}
\caption{Contributions of detector and thermal noise (gas collisions) for a $1$~s segment of the data shown in Section \ref{sec:HV_membrane_results}. (Top) Displacement spectral density showing the measured data and the sum of the noise contributions. The error bar centered at the peak of the gas collisions curve is calculated assuming a $5\%$ error in the radius of the sphere. (Bottom) Force spectral density showing that the thermal noise limited sensitivity coincides with the measured data on resonance to within the calculated error bar. The next most dominant source of noise is due to the detector electronics.}
\label{fig:noiseplot}
\end{figure}
where $\omega_0$ is the resonance angular frequency and $\Gamma$ is the damping rate. Figure \ref{fig:noiseplot} shows the displacement (top) and force (bottom) noise amplitude spectral densities including the raw experimental data, the detector noise with no sphere in the trap, theoretical calculations of the thermal noise (gas collisions), as well as the sum of these noise sources. Error bars are plotted at the peak frequency of the calculated thermal noise curves. This data shows, to within calculated error, that the limiting source of noise is gas collisions as opposed to other possible sources of technical noise such as fluctuating patch potentials or scattered light from the surface. This result is found for the particular case where we have chosen a spot over the membrane and mechanical degree of freedom which exhibits a minimal amount of surface induced force noise. The next leading source of noise is the electronic detector noise in the apparatus. 

The backaction noise spectrum for a given mechanical degree of freedom depends on the amount of collected scattered power $P_{s}$. Here we consider the $y$ degree of freedom which gives a backaction noise spectrum of \cite{tebben_optimaldetection}
\begin{equation}\label{eq:sffth}
    S_{ba}^y = \frac{2}{5}{\frac{\hbar k}{2\pi c} P_s}
\end{equation}
where $k$ is the wavenumber. For the given detection optics with NA = $0.24$ and laser power of $100$~mW, we estimate a collection efficiency of $\eta_c = 4$\%. Note that this does not fully account for the effect of the nearby surface on the dipole radiation pattern. A more rigorous calculation would require studying the dipole radiation pattern for a given orientation and distance from the surface using, for example, the Green dyadic method in the angular spectrum representation \cite{novotny_principles_2025}. With these values, we estimate a backaction noise of $5\times10^{-22}$~N/$\sqrt{\mathrm{Hz}}$.

Regarding the potential effects of SEFN, we measure the high vacuum force sensitivity for a sphere a distance of $1.6^{+1.9}_{-1.3}$~$\mu$m away from the thicker mirror surface with net charge of $|q| = 6e^-$ to be $2.2\pm 0.4$~aN$/\sqrt{\mathrm{Hz}}$ for the $z$ motion (see Fig. \ref{fig:chargecalibration}). The thermally limited force sensitivity for this sphere was calculated to be $2.0$~aN$/\sqrt{\mathrm{Hz}}$. Relating the power spectral density of the electric field noise and force noise via $S_{EE} = S_{FF}^{th}/q^2$, we can place an upper limit of $\sqrt{S_{EE}} \leq 2.3 \pm 0.4$~V/(m$\sqrt{\mathrm{Hz}}$) on the amount of SEFN assuming it were to account for all of the experimentally measured noise. In principle, the sensitivity to the electric field noise can be improved by increasing the sphere's net charge \textit{in-situ} \cite{siegel_optical_2025}. Given the thermal noise limited sensitivity for this data, this suggests that there is no significant SEFN to the level of sensitivity achieved by this apparatus. This enables the possibility of mapping out localized patches of SEFN over the conducting surface.

\section{Conclusion and Outlook} \label{sec:discuss}
We have demonstrated that optically levitated silica nanospheres can be used for three-dimensional force sensing while positioned at distances of order $1~\mu$m away from a Au coated SiN membrane. Operating in this way, the levitated silica sphere serves as a scanning probe microscope; the force sensitivity is orders of magnitude better than the other traditional force-sensitive scanning probe microscopes, such as conventional atomic force microscopes, which rely on a tethered microcantilever beam.  By employing three-dimensional laser feedback cooling of the center of mass motion of the levitated nanosphere, we have realized zeptonewton level force resolution in high vacuum while approximately $\sim 1~\mu$m away from the 370 nm thick metalized membrane.

Based on an analysis of the system noise, further improvements in force sensitivity at micron range from a metalized membrane appear promising. Currently, for on-resonance force sensing, the system is limited by intrinsic thermal force noise which can be reduced by achieving higher vacuum.  We anticipate that significant detector improvements are achievable pushing the sensitivity towards the shot noise limit. Crucially, we demonstrate that there are regions of the surface where the sensitivity is not fundamentally limited by noise sources due to the surface, which would pose a much larger technical roadblock to system improvement. Our results are promising for achieving zeptonewton resolution in micron-range gravity tests while still allowing for the presence of a conducting screening membrane.  A metalized membrane such as the one used in this work could also be considered for use as an electrostatic shield for future gravity mediated quantum entanglement experiments between spatially separated masses. 

Finally, we note that this experimental platform has potential applications to systematic studies of surface electric field noise at sub-$10$~$\mu$m distances in the $1-100$~kHz regime. Studying the spectrum of electric field noise in close proximity to conducting surfaces may have relevance for understanding the anomalous heating of the motion of ions near a surface due to high frequency electric field noise in surface-based ion traps \cite{RMPiontrapnoise,Abbasov2023,saarel_electrical_2026}.  Higher frequency studies of electric field noise at up to a few hundred kHz could be realized using a smaller laser beam waist or higher power as in Ref. \cite{montoya_scanning_2022,grinin_optically_2026}. 

\section{Acknowledgements} \label{sec:acknow}
We are grateful for helpful discussion with Eduardo Alejandro and Hartmut H\"affner. We acknowledge support from NSF grants PHY-2409472 and PHY-2111544, DARPA, the John Templeton Foundation, the W.M. Keck Foundation, the Gordon and Betty Moore Foundation Grant GBMF12328, DOI 10.37807/GBMF12328, the Alfred P. Sloan Foundation under Grant No. G-2023-21130

\section{Appendix A: Fiber interferometer and distance metrology} \label{sec:fiber_interferometer}

\begin{figure}
    \centering
    \includegraphics[width=\linewidth]{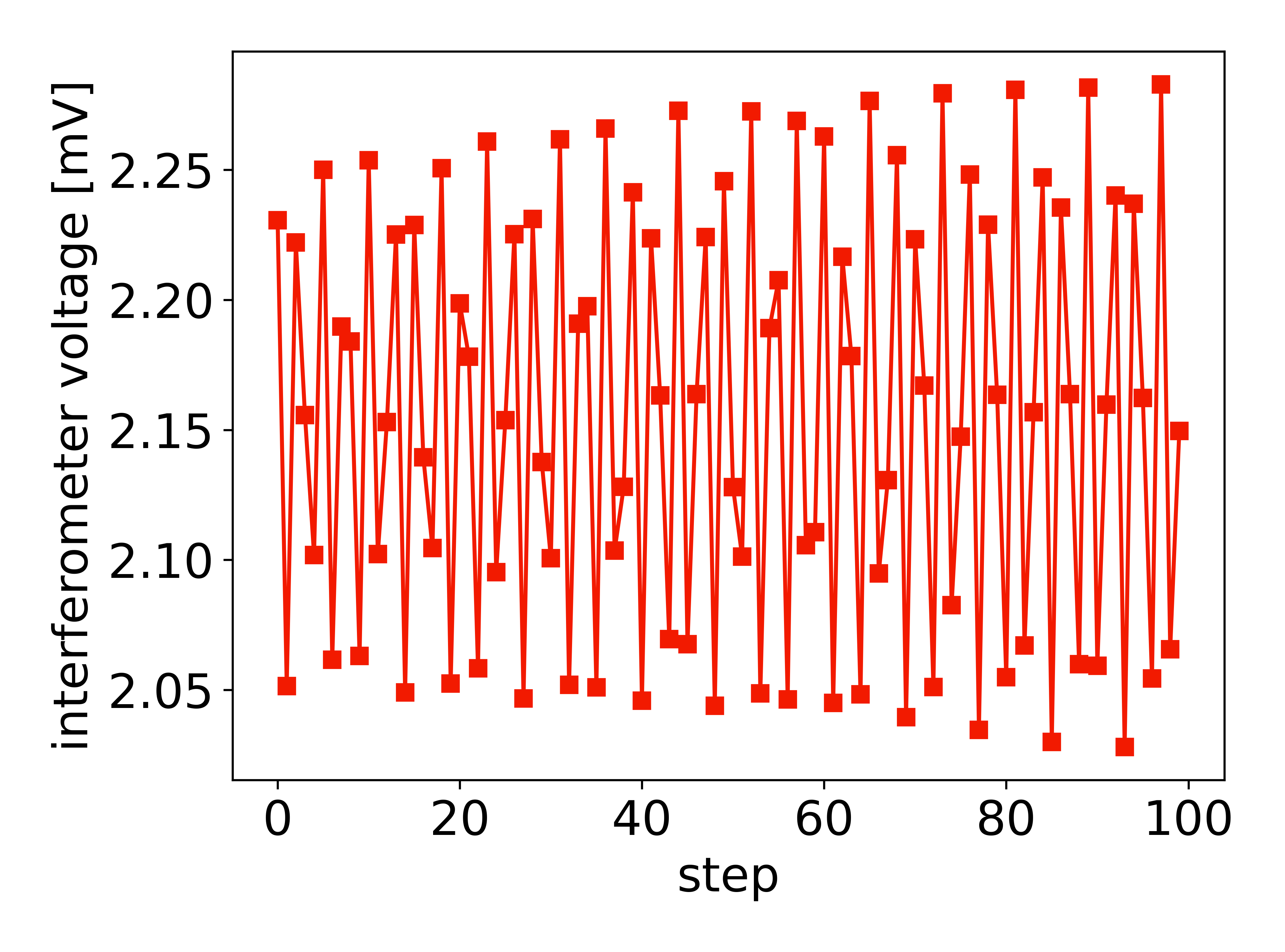}
    \caption{Fiber-based interferometer voltage signal on a photodetector as the dual-beam $z$ stage --- and the cleaved fiber face --- is moved in one step increments towards the knife-edged mirror holder. The spacing between fringes in the signal is equal to $327.5$~nm for our $1310$~nm light. }
    \label{fig:interferometer}
\end{figure}

In order to measure the distance traveled by the dual-beam stage along the optical axis $z$ we use a fiber-based interferometer built according to the design given in Ref. \cite{rugar_improved_1989}. We send 1310~nm light through a bare single-mode optical fiber that is attached to the 3D stage that holds the dual-beam collimator packages. A cleaved end of this fiber is pointed at a mirror affixed to the aluminum structure that holds the knife-edged gold-coated mirror (that is stationary along the $z$ direction); light that never exists the fiber but is instead reflected by the cleaved fiber face acts as the local oscillator for the interferometer. The interference pattern changes as the dual-beam stage moves along the $z$ direction and the distance between the fiber face and the mirror changes. We measure the interference by recording the voltage signal on a photodetector for every step the dual-beam $z$ stage takes, an example of which is shown in Fig. \ref{fig:interferometer}. The distance traveled by the stage can be found by phase unwinding the voltage signal. The fringes of the interferometer are spaced by $\lambda/4$, or $327.5$~nm for our $1310$~nm light. We use this fiber interferometer to position a sphere a known distance $z_0$ away from the mirror in the standing wave trap, which we discuss in the following paragraphs.

Prior to trapping a sphere, we fully insert the mirror into the counter-propagating beams and measure the foci offset $d$ by translating the dual-beam trap along the optical axis $z$ in $\sim$250~nm steps. We perform this foci offset measurement on the 370~nm thick membrane because its thickness is much less than the foci offset. We measure the power retro-reflected into the S- and P-beam paths with two separate photodetectors via upstream pickoffs (as shown in Fig. \ref{fig:experimentalsetup}a). The S- and P-beam foci are located where the retro-reflected powers are maximum, as determined by fitting the data. The location of the sphere’s trapping site is half-way ($d/2$) between these two foci locations. Using the fiber interferometer to count fringes and measure the stage's motion, we then position the dual-beam trap along $z$ so that the trapping location is in the same plane as the S-side of the mirror's surface. 

In order to account for a small angle between the optical axis $z$ and the mirror’s axis of motion $x$, we move the mirror along $x$ until we are barely on the knife-edge, as this is the first location on the mirror that the sphere sees when transitioning between traps. We then perform a measurement of the distance between the S-focus and the mirror’s surface, $l$. If the angle of the mirror were zero, we would recover $l=d/2$; however, for our measured angle of $1$~mrad, $l<d/2$ and the mirror is slightly titled towards the S-focus. We then re-position the mirror by moving along the trapping axis $z$ until $l=d/2$. Repeated measurements of $l-d/2$ after following the zeroing out procedure give a mean value of $0~\mu$m with a standard deviation of $0.7~\mu$m, where the uncertainty comes from the fit to the retro-reflected power data. Once the knife-edge of the mirror surface is at the trapping location, we move a known distance $z_0$ to position the trap for slicing a sphere. A similar experiment that used a mirror to transition between an optical tweezer and a retro-reflected standing wave trap showed that the sliced sphere’s position is very close to $z_0$ once transitioned into the standing wave trap, up to adjacent trapping sites \cite{montoya_scanning_2022}. We therefore report $z_0$ as our measurement of the sphere’s position. 

In addition to the leading uncertainty on the sphere's distance from the mirror due to the RMS motion of $1.8~\mu$m, we also quantify the uncertainty in the measurement of the location of the foci midpoint. By taking the standard deviation of repeated measurements of the distance $d/2$, we measure this uncertainty to be $0.3~\mu$m. The uncertainty on the sphere’s position is therefore the sum in quadrature of the uncertainty due to the RMS motion, measurement repeatability of the foci midpoint, and measurement uncertainty of the angle: the total uncertainty is given by $\delta_{z_0} = 1.9~\mu$m.

\section{Appendix B: Dual Beam Power balancing \label{sec:power_balancing}}
\begin{figure}[h!]
\label{fig:powerbalanceposition}
\includegraphics[width=\linewidth]{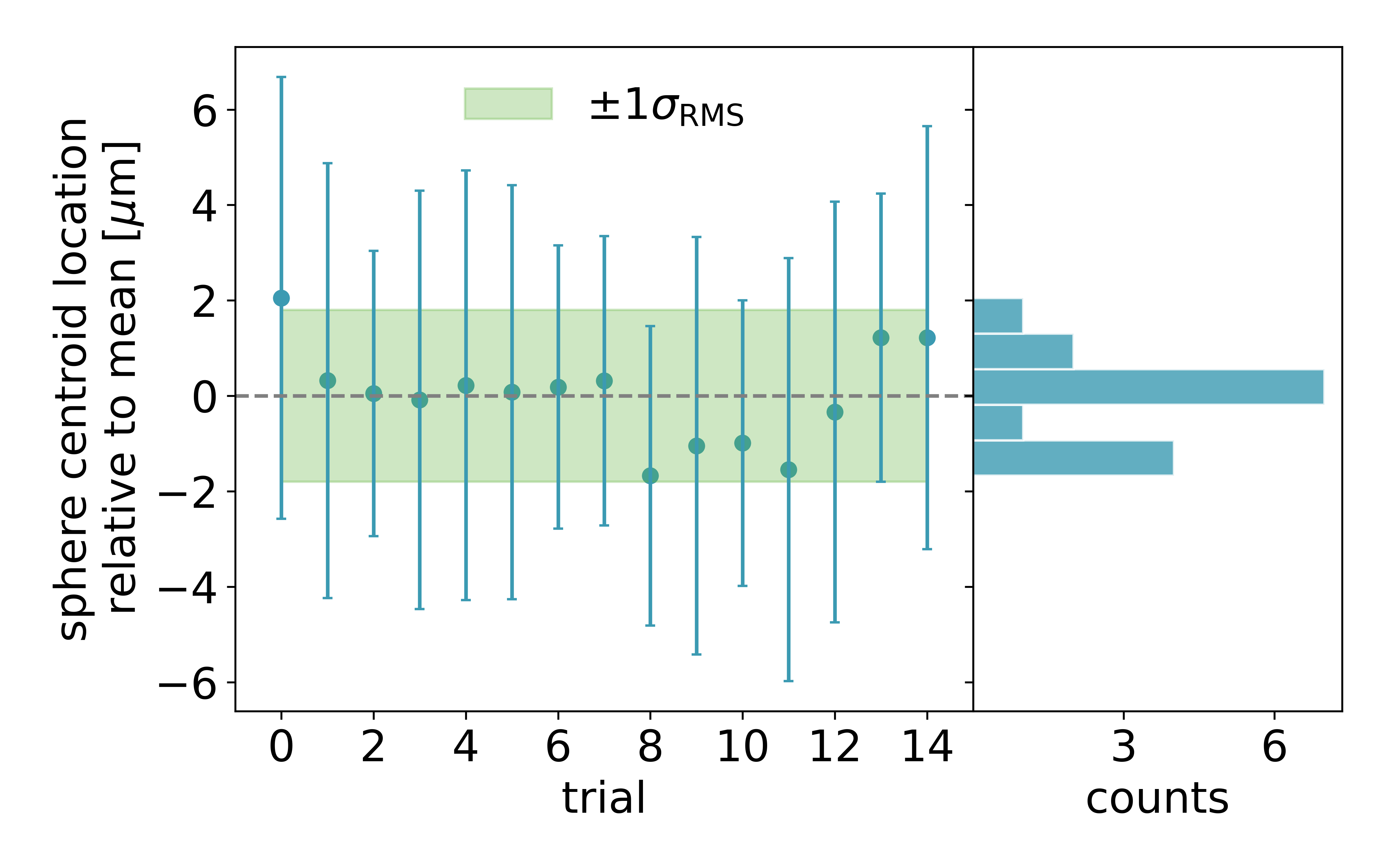}
\caption{Fifteen 300~nm silica spheres were imaged by a camera in the dual-beam trap after following the power balancing procedure prior to trapping. The centroids of the spheres' images are plotted relative to their mean in calibrated distance. Some uncertainty in the centroid location is expected due to the RMS motion of amplitude $1.8~\mu$m in the trap (shown in green shaded region in the plot). After the power balancing procedure, the spheres are all trapped at locations within the uncertainty of the RMS motion.}
\end{figure}

Our method for repeatable placement of the levitated sphere at single micron range from the conducting membrane relies on the assumption that the particle is initially trapped precisely halfway between the two foci of the initial dual-beam trap. To ensure the validity of this assumption, we determined the $\lambda/2$ wave-plate orientation before the PBS cube so that both the S- and P-beams have an equal optical power (measured to be $160$~mW using a power meter) at the trapping site in the vacuum chamber. We then monitored the cross-coupled laser power (i.e. the amount of power from the P-beam that couples into the S-beam collimator package and vice versa) between the two optical fibers as the vacuum was pumped down to $1$~mbar, which is our typical initial trapping pressure. We can therefore optimize the fiber coupling for each beam from outside the chamber to reproduce the cross-coupling values which gave equal power inside the chamber. Doing this cross-coupled optical power optimization each time before trapping a particle ensures the validity of the assumption that the particle is halfway between the two foci. We studied the variation in the trapping location for 15 trapped silica spheres after performing the cross-coupling optimization prior to trapping each sphere: we illuminated each sphere with $780$~nm light, took its image in a camera, and plotted its position by calibrating the camera's pixel-to-distance conversion. For each image, the sphere's pixel location was determined by performing a 2D Gaussian fit to the intensity; due to camera settings, the exposure was saturated which inflated the $\sigma_x, \sigma_y$ fit parameters. The pixel-to-distance calibration was done by moving the fiber-based dual-beam trap $20~\mu$m along the optical axis $z$ using the fiber interferometer while a sphere was trapped. The positions of the centroids of the 15 trapped spheres relative to their ensemble mean is shown in Fig. \ref{fig:powerbalanceposition}. This characterization shows that the power balancing method is repeatable up to the RMS displacement of the levitated particle in the dual beam trap which is calculated the be $z_{RMS} = 1.8$ $\mu$m. The imaging setup is constructed using a half-inch $25.4$~mm focal length lens inside the chamber which collimates the scattered light of the levitated particle out of the chamber. The imaging telescope is completed using a half-inch $500$~mm focal length lens magnifying the imaging of the particle by approximately a factor of $20$ onto a BFS-U3-16S2M-CS Blackfly Camera with $1.6$ megapixel resolution and $3.45$~$\mu$m pixel size.

\section{Appendix C: Internal temperature gradient of an optically trapped nanosphere}\label{sec:photophoresis}
Photophoresis is the phenomenon where small spheres, in the presence of a sufficient source of light, experience a force due to internal temperature gradients generated by the illumination. A prototypical example is Crookes radiometer \cite{crookes_xv_1874} where a set of vanes on a spindle are illuminated on one side by a light source. Considering a single vane, the illuminated side gets hot due to absorption while the other side is cool. Thus gas molecules hitting the hotter side scatter off with greater momentum than the cool side, generating a non-zero net force which turns the spindle. 

For the case of a silica nanosphere in thermal equilibrium with its environment, we consider the sum of the power absorbed from the laser $P_{abslaser}$ and black body radiation $P_{absbb}$ and equate it to the power lost due to gas collisions $P_{radcol}$ and emitted by the sphere via black body emission $P_{radbb}$. Considering the thermal radiation power for a Rayleigh sphere as well as the absorbed laser power and radiated power through gas collisions found in \cite{chang_cavity_2010}, we arrive at the set of equations
\begin{align}
    P_{absbb} &= \frac{72 \zeta(5) V}{\pi^2 c^3 \hbar^4}\IM\left[\frac{\epsilon_{bb}-1}{\epsilon_{bb}+2}\right](k_BT)^5\\
    P_{abslaser} &= \frac{12 \pi I_0 V}{\lambda} \IM\left[\frac{\epsilon_1+i \epsilon_2-1}{\epsilon_1 + i \epsilon_2+2}\right]
\end{align}
\begin{align}
    P_{radbb} &= \frac{72 \zeta(5) V}{\pi^2 c^3 \hbar^4}\IM\left[\frac{\epsilon_{bb}-1}{\epsilon_{bb}+2}\right](k_BT_{int})^5\\
    P_{radcol} &= -\alpha_g \sqrt{\frac{2}{3\pi}} \pi r^2 P v_{rms}\frac{\gamma_{sh}+1}{\gamma_{sh}-1}\left(\frac{T_{int}}{T}-1\right)
\end{align}
where $I_0$ is the laser intensity, $V$ is the sphere volume, $\lambda$ is the laser wavelength, c is the speed of light, $\hbar$ is the reduced Planck constant, $k_B$ is Boltzmann's constant, $T$ is the environmental temperature, $T_{int}$ is the internal temperature of the sphere, $\alpha_g$ is the accommodation coefficient, taken to be $0.25$, $P$ is the background gas presssure, $v_{rms}$ is the gas RMS velocity, $\gamma_{sh}$ is the gas specific heat ratio, $r$ is the sphere radius, $\epsilon_{bb}$ is the thermal emissivity of the material taken to be $0.1$, and $\epsilon_1$ and $\epsilon_2$ are the real and imaginary components of the complex permittivity, taken to be $2.1$ and $10^{-6}$, respectively.
We are specifically interested in the internal temperature gradient $\Gamma_i$ of the nanosphere. which for this case, is estimated as $\Gamma_i = \frac{T_0-T_r}{r}$ where $T_0$ is the temperature at the center of the sphere and $T_r$ is the temperature at the edge of the sphere. By equating the absorbed and radiated powers from above, these two temperatures are found by solving for $T_{int}$ for the case where the sphere is locally heated by the most intense part of the trapping beam (yielding $T_0$) and again for the case where the sphere is locally heated by the beam with the intensity one radius away from the waist in the direction orthogonal to the propagation axis of the laser (yielding $T_r$). With this estimate for the temperature gradient across the sphere, Eq. \ref{eq:Ft} can be used to calculate the force due to photophoresis across the Knudsen regimes as in Figure \ref{fig:phoreticforces}.

\section{Appendix D: Additional Scanning Force Microscopy with larger grids at larger distances}\label{sec:additionalscans}
\begin{figure}[h!]
\includegraphics[width=\linewidth]{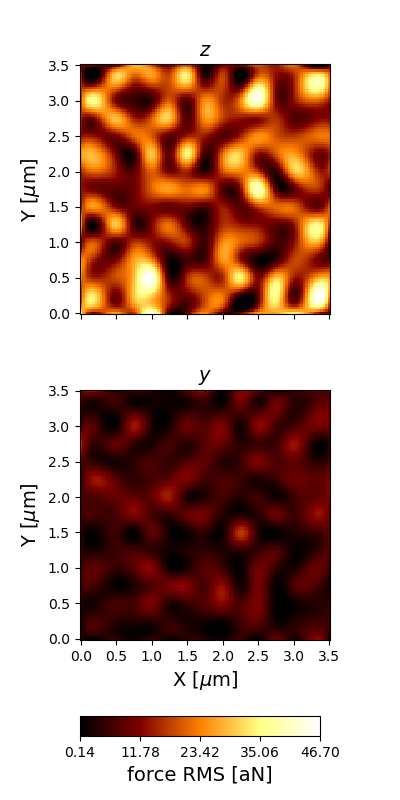}
\caption{RMS $z$ (top) and $y$ (bottom) force noise for a $15$ by $15$ step grid comprising of an approximately $3.5~\mu$m by $3.5~\mu$m patch of the gold coated surface. The measurement time at each grid point is $20$~s. The surface plot is made using a 2D cubic interpolation for visualization.}
\label{fig:fv_15x15}
\end{figure}

To demonstrate the modularity of this system as potentially useful for scanning force microscopy, we show an additional scan taken for a larger grid size and at a larger particle to surface distance. Figure \ref{fig:fv_15x15} presents force scanning data from two degrees of mechanical motion over a $3.5 \times 3.5$~$\mu$m grid taken with $250$~nm sized steps. Twenty seconds of data was taken at each grid point, and the particle to surface distance was measured to be approximately $11$~$\mu$m.  For each of these grids, after $20$~s of measurement time, the mean force in the $z$ direction was $17$~aN and $5.8$~aN in the $y$ direction. The estimated thermal-noise-limited minimum detectable force with $20$~s of measurement time for the $z$ direction is $6.9$~aN and $2.5$~aN for the $y$ direction.

\section{Appendix E: Back reflected light measurements}
\label{sec:backreflectedpower}
\begin{figure}
\includegraphics[width=\linewidth]{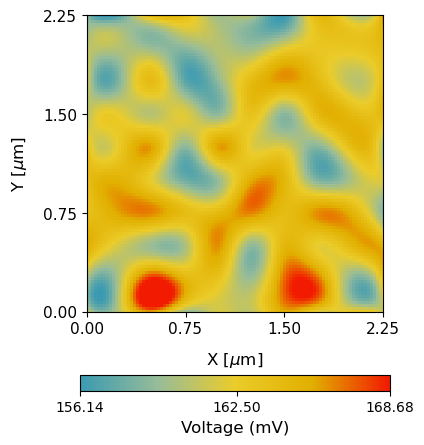}
\caption{Photodetector voltage of the back reflected trapping laser power taken in tandem with the data shown in Figure \ref{fig:membranegrid}. A 2D cubic
spline interpolation between grid points was used for data visualization.}
\label{fig:backreflectedpower}
\end{figure}

While scanning the particle over patches of the surface, the DC value of the back reflected laser is measured at each grid point to study possible noise due to light scattering. We assume that if light scattering from defects or contaminants on the surface were a dominant source of noise, we should see some correlation between changes in force noise and the back reflected light over the grid. The back reflected light is coupled back through the optical fiber which carries the S-beam, and the DC voltage is measured with a photodetector placed outside the vacuum chamber.  This data is shown in Figure \ref{fig:backreflectedpower}.  
The average voltage over this grid is $162.4$~mV with a standard deviation of $2.7$~mV. 
Point by point correlation coefficients between this data and each of the force noise grids in Figure \ref{fig:membranegrid} are calculated and yield values of -0.13, 0.17, and -0.13 for the x, y and z degrees of freedom respectively. Importantly, the alignment of the photodetector remains constant for the duration of the scanning measurement. These values indicate that the changes in the photodetector voltage are not correlated with changes in force noise for any of the particle's mechanical degrees of freedom. Thus we infer that scattered light off of surface defects and other contaminants, which would change the amount of back reflected light, is not a primary source of force noise over this patch of the metalized membrane.  

\section{Appendix F: Lateral and Axial Resolution}
\label{sec:resolution}
Here we try to estimate the lateral and axial spatial resolution of this particle-near-conducting surface system as it relates to its possible use in studying localized patches of SEFN which are resonant with the mechanical motion of the levitated sphere. We do not consider topographical mapping as changes in the surface's structure could also affect the trap potential landscape. In the case of the axial direction, we can calculate the thermal noise limited displacement resolution on resonance via $z_{min} = \frac{F_{min}}{k} Q$. Assuming a $1$~s measurement time, and using the values for the sphere studied in Section \ref{sec:force_and_charge_calibration}, we calculate a value of $z_{min} \approx 30$~pm.

To estimate the lateral resolution for the case of a charged sphere scanned over a conducting surface, we invoke the method of images \cite{zangwill_modern_2013} to find the induced surface charge density
\begin{equation}
    \sigma = -\frac{q}{2\pi}\frac{z_0}{(r^2 + z_0^2)^{3/2}}
\end{equation}
where $q$ is the charge of the electron, $r$ is the radial distance on the flat conducting surface and $z_0$ is the vertical distance of the charged particle above the surface. This estimation treats the levitated sphere as a point charge and does not account for dipole or patch potential effects. By defining the lateral resolution as the full width at half maximum of this surface charge density, we characterize the effective area of the surface over which the charged particle is sensitive to changes in electric forces. Half of the maximum value of the surface charge density is $\frac{1}{2}\sigma (r = 0) = -\frac{q}{4\pi}\frac{1}{z_0^2}$, and the full width ($2r$) at this value is $1.53 z_0$.
\bibliographystyle{apsrev4-2}
\bibliography{Sections/References}

@article{Mestres2015,
author = {Mestres, Pau and Berthelot, Johann and Spasenovi{\'{c}}, Marko and Gieseler, Jan and Novotny, Lukas and Quidant, Romain},
doi = {10.1063/1.4933180},
issn = {0003-6951},
journal = {Applied Physics Letters},
month = {oct},
number = {15},
pages = {151102},
title = {{Cooling and manipulation of a levitated nanoparticle with an optical fiber trap}},
url = {http://aip.scitation.org/doi/10.1063/1.4933180},
volume = {107},
year = {2015}
}

@article{Kiesel2013,
author = {Kiesel, Nikolai and Blaser, Florian and Deli{\'{c}}, Uro{\v{s}} and Grass, David and Kaltenbaek, Rainer and Aspelmeyer, Markus},
doi = {10.1073/pnas.1309167110},
issn = {00278424},
journal = {Proceedings of the National Academy of Sciences of the United States of America},
month = {aug},
number = {35},
pages = {14180--14185},
pmid = {23940352},
publisher = {National Academy of Sciences},
title = {{Cavity cooling of an optically levitated submicron particle}},
url = {https://www.pnas.org/content/110/35/14180},
volume = {110},
year = {2013}
}

@article{Millen2014,
archivePrefix = {arXiv},
arxivId = {1309.3990},
author = {Millen, J. and Deesuwan, T. and Barker, P. and Anders, J.},
doi = {10.1038/nnano.2014.82},
eprint = {1309.3990},
issn = {17483395},
journal = {Nature Nanotechnology},
number = {6},
pages = {425--429},
publisher = {Nature Publishing Group},
title = {{Nanoscale temperature measurements using non-equilibrium Brownian dynamics of a levitated nanosphere}},
volume = {9},
year = {2014}
}

@article{Price2015,
author = {Price, C. J. and Donnelly, T. D. and Giltrap, S. and Stuart, N. H. and Parker, S. and Patankar, S. and Lowe, H. F. and Drew, D. and Gumbrell, E. T. and Smith, R. A.},
doi = {10.1063/1.4908285},
issn = {10897623},
journal = {Review of Scientific Instruments},
number = {3},
title = {{An in-vacuo optical levitation trap for high-intensity laser interaction experiments with isolated microtargets}},
url = {http://dx.doi.org/10.1063/1.4908285},
volume = {86},
year = {2015}
}

@article{ranjit_zeptonewton_2016,
	title = {Zeptonewton force sensing with nanospheres in an optical lattice},
	volume = {93},
	copyright = {http://link.aps.org/licenses/aps-default-license},
	issn = {2469-9926, 2469-9934},
	url = {https://link.aps.org/doi/10.1103/PhysRevA.93.053801},
	doi = {10.1103/PhysRevA.93.053801},
	language = {english},
	number = {5},
	urldate = {2026-07-09},
	journal = {Physical Review A},
	author = {Ranjit, Gambhir and Cunningham, Mark and Casey, Kirsten and Geraci, Andrew A.},
	month = may,
	year = {2016},
	pages = {053801},
}

@article{hempston_force_2017,
	title = {Force sensing with an optically levitated charged nanoparticle},
	volume = {111},
	issn = {0003-6951, 1077-3118},
	url = {https://pubs.aip.org/apl/article/111/13/133111/34032/Force-sensing-with-an-optically-levitated-charged},
	doi = {10.1063/1.4993555},
	language = {english},
	number = {13},
	urldate = {2026-07-09},
	journal = {Applied Physics Letters},
	author = {Hempston, David and Vovrosh, Jamie and Toroš, Marko and Winstone, George and Rashid, Muddassar and Ulbricht, Hendrik},
	month = sep,
	year = {2017},
	pages = {133111},
}

@article{kawasaki_high_2020,
	title = {High sensitivity, levitated microsphere apparatus for short-distance force measurements},
	volume = {91},
	issn = {0034-6748, 1089-7623},
	url = {https://pubs.aip.org/rsi/article/91/8/083201/989394/High-sensitivity-levitated-microsphere-apparatus},
	doi = {10.1063/5.0011759},
	language = {english},
	number = {8},
	urldate = {2026-07-09},
	journal = {Review of Scientific Instruments},
	author = {Kawasaki, Akio and Fieguth, Alexander and Priel, Nadav and Blakemore, Charles P. and Martin, Denzal and Gratta, Giorgio},
	month = aug,
	year = {2020},
	pages = {083201},
}

@article{delic_cooling_2020,
	title = {Cooling of a levitated nanoparticle to the motional quantum ground state},
	volume = {367},
	issn = {0036-8075, 1095-9203},
	url = {https://www.science.org/doi/10.1126/science.aba3993},
	doi = {10.1126/science.aba3993},
	language = {english},
	number = {6480},
	urldate = {2026-07-09},
	journal = {Science},
	author = {Delić, Uroš and Reisenbauer, Manuel and Dare, Kahan and Grass, David and Vuletić, Vladan and Kiesel, Nikolai and Aspelmeyer, Markus},
	month = feb,
	year = {2020},
	pages = {892--895},
}

@article{monteiro_force_2020,
	title = {Force and acceleration sensing with optically levitated nanogram masses at microkelvin temperatures},
	volume = {101},
	issn = {2469-9926, 2469-9934},
	url = {https://link.aps.org/doi/10.1103/PhysRevA.101.053835},
	doi = {10.1103/PhysRevA.101.053835},
	language = {english},
	number = {5},
	urldate = {2026-07-09},
	journal = {Physical Review A},
	author = {Monteiro, Fernando and Li, Wenqiang and Afek, Gadi and Li, Chang-ling and Mossman, Michael and Moore, David C.},
	month = may,
	year = {2020},
	pages = {053835},
}

@article{li_millikelvin_2011,
	title = {Millikelvin cooling of an optically trapped microsphere in vacuum},
	volume = {7},
	issn = {1745-2473, 1745-2481},
	url = {https://www.nature.com/articles/nphys1952},
	doi = {10.1038/nphys1952},
	language = {english},
	number = {7},
	urldate = {2026-07-09},
	journal = {Nature Physics},
	author = {Li, Tongcang and Kheifets, Simon and Raizen, Mark G.},
	month = jul,
	year = {2011},
	pages = {527--530},
}

@article{grinin_optically_2026,
	title = {Optically trapping nanospheres at micrometer range from a tilted mirror},
	volume = {25},
	issn = {2331-7019},
	url = {https://link.aps.org/doi/10.1103/nx9n-ypxl},
	doi = {10.1103/nx9n-ypxl},
	language = {english},
	number = {2},
	urldate = {2026-07-09},
	journal = {Physical Review Applied},
	author = {Grinin, Alexey and Dana, Andrew and Nguyen, Mark and Alejandro, Eduardo and Geraci, Andrew A.},
	month = feb,
	year = {2026},
	pages = {024067},
}

@article{montoya_scanning_2022,
	title = {Scanning force sensing at micrometer distances from a conductive surface with nanospheres in an optical lattice},
	volume = {61},
	issn = {1559-128X, 2155-3165},
	url = {https://opg.optica.org/abstract.cfm?URI=ao-61-12-3486},
	doi = {10.1364/AO.457148},
	language = {english},
	number = {12},
	urldate = {2026-07-09},
	journal = {Applied Optics},
	author = {Montoya, Cris and Alejandro, Eduardo and Eom, William and Grass, Daniel and Clarisse, Nicolas and Witherspoon, Apryl and Geraci, Andrew A.},
	month = apr,
	year = {2022},
	pages = {3486},
}

@article{diehl_optical_2018,
	title = {Optical levitation and feedback cooling of a nanoparticle at subwavelength distances from a membrane},
	volume = {98},
	issn = {2469-9926, 2469-9934},
	url = {https://link.aps.org/doi/10.1103/PhysRevA.98.013851},
	doi = {10.1103/PhysRevA.98.013851},
	language = {english},
	number = {1},
	urldate = {2026-07-09},
	journal = {Physical Review A},
	author = {Diehl, Rozenn and Hebestreit, Erik and Reimann, René and Tebbenjohanns, Felix and Frimmer, Martin and Novotny, Lukas},
	month = jul,
	year = {2018},
	pages = {013851},
}

@article{winstone_direct_2018,
	title = {Direct measurement of the electrostatic image force of a levitated charged nanoparticle close to a surface},
	volume = {98},
	issn = {2469-9926, 2469-9934},
	url = {https://link.aps.org/doi/10.1103/PhysRevA.98.053831},
	doi = {10.1103/PhysRevA.98.053831},
	language = {english},
	number = {5},
	urldate = {2026-07-09},
	journal = {Physical Review A},
	author = {Winstone, George and Bennett, Robert and Rademacher, Markus and Rashid, Muddassar and Buhmann, Stefan and Ulbricht, Hendrik},
	month = nov,
	year = {2018},
	pages = {053831},
}

@article{bose_spin_2017,
	title = {Spin {Entanglement} {Witness} for {Quantum} {Gravity}},
	volume = {119},
	copyright = {https://link.aps.org/licenses/aps-default-license},
	issn = {0031-9007, 1079-7114},
	url = {https://link.aps.org/doi/10.1103/PhysRevLett.119.240401},
	doi = {10.1103/PhysRevLett.119.240401},
	language = {english},
	number = {24},
	urldate = {2026-07-09},
	journal = {Physical Review Letters},
	author = {Bose, Sougato and Mazumdar, Anupam and Morley, Gavin W. and Ulbricht, Hendrik and Toroš, Marko and Paternostro, Mauro and Geraci, Andrew A. and Barker, Peter F. and Kim, M. S. and Milburn, Gerard},
	month = dec,
	year = {2017},
	pages = {240401},
}

@article{marletto_gravitationally_2017,
	title = {Gravitationally {Induced} {Entanglement} between {Two} {Massive} {Particles} is {Sufficient} {Evidence} of {Quantum} {Effects} in {Gravity}},
	volume = {119},
	copyright = {https://link.aps.org/licenses/aps-default-license},
	issn = {0031-9007, 1079-7114},
	url = {https://link.aps.org/doi/10.1103/PhysRevLett.119.240402},
	doi = {10.1103/PhysRevLett.119.240402},
	language = {english},
	number = {24},
	urldate = {2026-07-09},
	journal = {Physical Review Letters},
	author = {Marletto, C. and Vedral, V.},
	month = dec,
	year = {2017},
	pages = {240402},
}

@article{frangeskou_pure_2018,
	title = {Pure nanodiamonds for levitated optomechanics in vacuum},
	volume = {20},
	issn = {1367-2630},
	url = {https://iopscience.iop.org/article/10.1088/1367-2630/aab700},
	doi = {10.1088/1367-2630/aab700},
	language = {english},
	number = {4},
	urldate = {2026-07-16},
	journal = {New Journal of Physics},
	author = {Frangeskou, A C and Rahman, A T M A and Gines, L and Mandal, S and Williams, O A and Barker, P F and Morley, G W},
	month = apr,
	year = {2018},
	pages = {043016},
}

@article{tebbenjohanns_quantum_2021,
	title = {Quantum control of a nanoparticle optically levitated in cryogenic free space},
	volume = {595},
	issn = {0028-0836, 1476-4687},
	url = {https://www.nature.com/articles/s41586-021-03617-w},
	doi = {10.1038/s41586-021-03617-w},
	language = {english},
	number = {7867},
	urldate = {2026-07-17},
	journal = {Nature},
	author = {Tebbenjohanns, Felix and Mattana, M. Luisa and Rossi, Massimiliano and Frimmer, Martin and Novotny, Lukas},
	month = jul,
	year = {2021},
	pages = {378--382},
}

@article{kamba_quantum_2025,
	title = {Quantum squeezing of a levitated nanomechanical oscillator},
	volume = {389},
	issn = {0036-8075, 1095-9203},
	url = {https://www.science.org/doi/10.1126/science.ady4652},
	doi = {10.1126/science.ady4652},
	language = {english},
	number = {6766},
	urldate = {2026-07-17},
	journal = {Science},
	author = {Kamba, Mitsuyoshi and Hara, Naoki and Aikawa, Kiyotaka},
	month = sep,
	year = {2025},
	pages = {1225--1228},
}

@article{blakemore_three-dimensional_2019,
	title = {Three-dimensional force-field microscopy with optically levitated microspheres},
	volume = {99},
	issn = {2469-9926, 2469-9934},
	url = {https://link.aps.org/doi/10.1103/PhysRevA.99.023816},
	doi = {10.1103/PhysRevA.99.023816},
	language = {english},
	number = {2},
	urldate = {2026-07-17},
	journal = {Physical Review A},
	author = {Blakemore, Charles P. and Rider, Alexander D. and Roy, Sandip and Wang, Qidong and Kawasaki, Akio and Gratta, Giorgio},
	month = feb,
	year = {2019},
	pages = {023816},
}

@article{liang_yoctonewton_2023,
	title = {Yoctonewton force detection based on optically levitated oscillator},
	volume = {3},
	issn = {26673258},
	url = {https://linkinghub.elsevier.com/retrieve/pii/S2667325822003879},
	doi = {10.1016/j.fmre.2022.09.021},
	language = {english},
	number = {1},
	urldate = {2026-07-17},
	journal = {Fundamental Research},
	author = {Liang, Tao and Zhu, Shaochong and He, Peitong and Chen, Zhiming and Wang, Yingying and Li, Cuihong and Fu, Zhenhai and Gao, Xiaowen and Chen, Xinfan and Li, Nan and Zhu, Qi and Hu, Huizhu},
	month = jan,
	year = {2023},
	pages = {57--62},
}

@article{rahman_laser_2017,
	title = {Laser refrigeration, alignment and rotation of levitated {Yb3}+:{YLF} nanocrystals},
	volume = {11},
	issn = {1749-4885, 1749-4893},
	shorttitle = {Laser refrigeration, alignment and rotation of levitated {Yb3}+},
	url = {https://www.nature.com/articles/s41566-017-0005-3},
	doi = {10.1038/s41566-017-0005-3},
	language = {english},
	number = {10},
	urldate = {2026-07-16},
	journal = {Nature Photonics},
	author = {Rahman, A. T. M. Anishur and Barker, P. F.},
	month = oct,
	year = {2017},
	pages = {634--638},
}

@article{schut_micrometer-size_2024,
	title = {Micrometer-size spatial superpositions for the {QGEM} protocol via screening and trapping},
	volume = {6},
	issn = {2643-1564},
	url = {https://link.aps.org/doi/10.1103/PhysRevResearch.6.013199},
	doi = {10.1103/PhysRevResearch.6.013199},
	language = {english},
	number = {1},
	urldate = {2026-07-09},
	journal = {Physical Review Research},
	author = {Schut, Martine and Geraci, Andrew and Bose, Sougato and Mazumdar, Anupam},
	month = feb,
	year = {2024},
	pages = {013199},
}

@article{schut_relaxation_2023,
	title = {Relaxation of experimental parameters in a quantum-gravity-induced entanglement of masses protocol using electromagnetic screening},
	volume = {5},
	issn = {2643-1564},
	url = {https://link.aps.org/doi/10.1103/PhysRevResearch.5.043170},
	doi = {10.1103/PhysRevResearch.5.043170},
	language = {english},
	number = {4},
	urldate = {2026-07-09},
	journal = {Physical Review Research},
	author = {Schut, Martine and Grinin, Alexey and Dana, Andrew and Bose, Sougato and Geraci, Andrew and Mazumdar, Anupam},
	month = nov,
	year = {2023},
	pages = {043170},
}

@article{chang_cavity_2010,
	title = {Cavity opto-mechanics using an optically levitated nanosphere},
	volume = {107},
	issn = {0027-8424, 1091-6490},
	url = {https://pnas.org/doi/full/10.1073/pnas.0912969107},
	doi = {10.1073/pnas.0912969107},
	language = {english},
	number = {3},
	urldate = {2026-07-09},
	journal = {Proceedings of the National Academy of Sciences},
	author = {Chang, D. E. and Regal, C. A. and Papp, S. B. and Wilson, D. J. and Ye, J. and Painter, O. and Kimble, H. J. and Zoller, P.},
	month = jan,
	year = {2010},
	pages = {1005--1010},
}

@article{jakubec_decoherence_2025,
	title = {Decoherence and {Brownian} motion of a polarizable particle near a medium},
	volume = {112},
	issn = {2469-9926, 2469-9934},
	url = {https://link.aps.org/doi/10.1103/6m4t-jm4x},
	doi = {10.1103/6m4t-jm4x},
	language = {english},
	number = {4},
	urldate = {2026-07-09},
	journal = {Physical Review A},
	author = {Jakubec, Clemens and Jarzynski, Christopher and Sinha, Kanu},
	month = oct,
	year = {2025},
	pages = {042225},
}

@article{winstone_optical_2022,
	title = {Optical {Trapping} of {High}-{Aspect}-{Ratio} {NaYF} {Hexagonal} {Prisms} for {kHz}-{MHz} {Gravitational} {Wave} {Detectors}},
	volume = {129},
	issn = {0031-9007, 1079-7114},
	url = {https://link.aps.org/doi/10.1103/PhysRevLett.129.053604},
	doi = {10.1103/PhysRevLett.129.053604},
	language = {english},
	number = {5},
	urldate = {2026-07-09},
	journal = {Physical Review Letters},
	author = {Winstone, George and Wang, Zhiyuan and Klomp, Shelby and Felsted, Robert G. and Laeuger, Andrew and Gupta, Chaman and Grass, Daniel and Aggarwal, Nancy and Sprague, Jacob and Pauzauskie, Peter J. and Larson, Shane L. and Kalogera, Vicky and Geraci, Andrew A. and {LSD Collaboration}},
	month = jul,
	year = {2022},
	pages = {053604},
}

@article{grinin_localized_2026,
	title = {Localized efficient in-vacuum loading of ∼0.1–10 \textit{μ} m spherical and plate-like particles into optical traps using a pulled glass capillary},
	volume = {97},
	issn = {0034-6748, 1089-7623},
	url = {https://pubs.aip.org/rsi/article/97/9/093204/3405001/Localized-efficient-in-vacuum-loading-of-0-1-10-m},
	doi = {10.1063/5.0342622},
	language = {english},
	number = {9},
	urldate = {2026-09-23},
	journal = {Review of Scientific Instruments},
	author = {Grinin, Alexey and Dana, Andrew and Nguyen, Mark and Grudichak, Scott and Guy, Katarina Boskovic and Klomp, Shelby and Elahi, Shafaq Gulzar and Borden, Sam and Wang, Zhiyuan and Winstone, George and Geraci, Andrew A.},
	month = sep,
	year = {2026},
	pages = {093204},
}

@article{geraci_short-range_2010,
	title = {Short-{Range} {Force} {Detection} {Using} {Optically} {Cooled} {Levitated} {Microspheres}},
	volume = {105},
	copyright = {http://link.aps.org/licenses/aps-default-license},
	issn = {0031-9007, 1079-7114},
	url = {https://link.aps.org/doi/10.1103/PhysRevLett.105.101101},
	doi = {10.1103/PhysRevLett.105.101101},
	language = {english},
	number = {10},
	urldate = {2026-07-09},
	journal = {Physical Review Letters},
	author = {Geraci, Andrew A. and Papp, Scott B. and Kitching, John},
	month = aug,
	year = {2010},
	pages = {101101},
}

@article{piotrowski_simultaneous_2023,
	title = {Simultaneous ground-state cooling of two mechanical modes of a levitated nanoparticle},
	volume = {19},
	issn = {1745-2473, 1745-2481},
	url = {https://www.nature.com/articles/s41567-023-01956-1},
	doi = {10.1038/s41567-023-01956-1},
	language = {english},
	number = {7},
	urldate = {2026-07-09},
	journal = {Nature Physics},
	author = {Piotrowski, Johannes and Windey, Dominik and Vijayan, Jayadev and Gonzalez-Ballestero, Carlos and De Los Ríos Sommer, Andrés and Meyer, Nadine and Quidant, Romain and Romero-Isart, Oriol and Reimann, René and Novotny, Lukas},
	month = jul,
	year = {2023},
	pages = {1009--1013},
}

@article{ranfagni_two-dimensional_2022,
	title = {Two-dimensional quantum motion of a levitated nanosphere},
	volume = {4},
	issn = {2643-1564},
	url = {https://link.aps.org/doi/10.1103/PhysRevResearch.4.033051},
	doi = {10.1103/PhysRevResearch.4.033051},
	language = {english},
	number = {3},
	urldate = {2026-07-09},
	journal = {Physical Review Research},
	author = {Ranfagni, A. and Børkje, K. and Marino, F. and Marin, F.},
	month = jul,
	year = {2022},
	pages = {033051},
}

@book{braginskyweakforces,
  author       = {Braginsky, V B and Manukin, A B},
  title        = {Measurement of weak forces in physics experiments},
  url          = {https://www.osti.gov/biblio/6644268},
  place        = {United States},
  publisher    = {University of Chicago Press,Chicago},
  year         = {1977},
  month        = {01}}

@article{ranjit_attonewton_2015,
	title = {Attonewton force detection using microspheres in a dual-beam optical trap in high vacuum},
	volume = {91},
	copyright = {http://link.aps.org/licenses/aps-default-license},
	issn = {1050-2947, 1094-1622},
	url = {https://link.aps.org/doi/10.1103/PhysRevA.91.051805},
	doi = {10.1103/PhysRevA.91.051805},
	language = {english},
	number = {5},
	urldate = {2026-07-16},
	journal = {Physical Review A},
	author = {Ranjit, Gambhir and Atherton, David P. and Stutz, Jordan H. and Cunningham, Mark and Geraci, Andrew A.},
	month = may,
	year = {2015},
	pages = {051805},
}

@article{riviere_thermometry_2022,
	title = {Thermometry of an optically levitated nanodiamond},
	volume = {4},
	issn = {2639-0213},
	url = {https://pubs.aip.org/aqs/article/4/3/030801/2835258/Thermometry-of-an-optically-levitated-nanodiamond},
	doi = {10.1116/5.0093600},
	language = {english},
	number = {3},
	urldate = {2026-07-16},
	journal = {AVS Quantum Science},
	author = {Rivière, François and De Guillebon, Timothée and Maumet, Léo and Hétet, Gabriel and Schmidt, Martin and Lauret, Jean-Sébastien and Rondin, Loïc},
	month = sep,
	year = {2022},
	pages = {030801},
}

@article{millen_nanoscale_2014,
	title = {Nanoscale temperature measurements using non-equilibrium {Brownian} dynamics of a levitated nanosphere},
	volume = {9},
	issn = {1748-3387, 1748-3395},
	url = {https://www.nature.com/articles/nnano.2014.82},
	doi = {10.1038/nnano.2014.82},
	language = {english},
	number = {6},
	urldate = {2026-07-16},
	journal = {Nature Nanotechnology},
	author = {Millen, J. and Deesuwan, T. and Barker, P. and Anders, J.},
	month = jun,
	year = {2014},
	pages = {425--429},
}

@article{zhang_determining_2023,
	title = {Determining the internal temperature of an optically levitated nanoparticle in vacuum by doped- {Er} 3 + -ion luminescence},
	volume = {108},
	issn = {2469-9926, 2469-9934},
	url = {https://link.aps.org/doi/10.1103/PhysRevA.108.033503},
	doi = {10.1103/PhysRevA.108.033503},
	language = {english},
	number = {3},
	urldate = {2026-07-16},
	journal = {Physical Review A},
	author = {Zhang, Baobao and Guo, Xiaojun and Yu, Xudong and Xiao, Yanzhen and Fu, Zhengkun and Zhang, Zhenglong and Zheng, Hairong},
	month = sep,
	year = {2023},
	pages = {033503},
}

@article{tseng_search_2025,
	title = {Search for {Dark} {Matter} {Scattering} from {Optically} {Levitated} {Nanoparticles}},
	volume = {6},
	issn = {2691-3399},
	url = {https://link.aps.org/doi/10.1103/j76m-gcp1},
	doi = {10.1103/j76m-gcp1},
	language = {english},
	number = {4},
	urldate = {2026-07-16},
	journal = {PRX Quantum},
	author = {Tseng, Yu-Han and Penny, T.W. and Siegel, Benjamin and Wang, Jiaxiang and Moore, David C.},
	month = dec,
	year = {2025},
	pages = {040367},
}

@misc{saarel_electrical_2026,
	title = {Electrical {Noise} {Produced} by {Micron}-{Sized} {Particles} above a {Surface} {Paul} {Trap}},
	url = {http://arxiv.org/abs/2606.19585},
	doi = {10.48550/arXiv.2606.19585},
	language = {english},
	urldate = {2026-07-15},
	publisher = {arXiv},
	author = {Saarel, Ben and Sahin, Ozgur and N'Diaye, Alpha T. and Häffner, Hartmut},
	month = jun,
	year = {2026},
	note = {arXiv:2606.19585 [quant-ph]},
}

@article{blakemore_search_2021,
	title = {Search for non-{Newtonian} interactions at micrometer scale with a levitated test mass},
	volume = {104},
	issn = {2470-0010, 2470-0029},
	url = {https://link.aps.org/doi/10.1103/PhysRevD.104.L061101},
	doi = {10.1103/PhysRevD.104.L061101},
	language = {english},
	number = {6},
	urldate = {2026-07-18},
	journal = {Physical Review D},
	author = {Blakemore, Charles P. and Fieguth, Alexander and Kawasaki, Akio and Priel, Nadav and Martin, Denzal and Rider, Alexander D. and Wang, Qidong and Gratta, Giorgio},
	month = sep,
	year = {2021},
	pages = {L061101},
}

@book{zangwill_modern_2013,
	address = {Cambridge},
	title = {Modern electrodynamics},
	isbn = {978-0-521-89697-9},
	language = {english},
	publisher = {Cambridge university press},
	author = {Zangwill, Andrew},
	year = {2013},
}

@article{neuhaus_python_2024,
	title = {Python {Red} {Pitaya} {Lockbox} ({PyRPL}): {An} open source software package for digital feedback control in quantum optics experiments},
	volume = {95},
	issn = {0034-6748},
	shorttitle = {Python {Red} {Pitaya} {Lockbox} ({PyRPL})},
	url = {https://doi.org/10.1063/5.0178481},
	doi = {10.1063/5.0178481},
	number = {3},
	urldate = {2026-09-14},
	journal = {Review of Scientific Instruments},
	author = {Neuhaus, Leonhard and Croquette, Michaël and Metzdorff, Rémi and Chua, Sheon and Jacquet, Pierre-Edouard and Journeaux, Alexandre and Heidmann, Antoine and Briant, Tristan and Jacqmin, Thibaut and Cohadon, Pierre-François and Deléglise, Samuel},
	month = mar,
	year = {2024},
	pages = {033003},

}

@misc{harvey_nanomechanical_2022,
	title = {Nanomechanical testing of silica nanospheres for levitated optomechanics experiments},
	url = {https://arxiv.org/abs/2208.02032v1},
	language = {english},
	urldate = {2026-09-15},
	journal = {arXiv.org},
	author = {Harvey, Cayla R. and Weisman, Evan and Galla, Chethn and Danenberg, Ryan and Hu, Qiyuan and Singh, Swati and Geraci, Andrew A. and Pathak, Siddhartha},
	month = aug,
	year = {2022},
}

@book{novotny_principles_2025,
	edition = {3},
	title = {Principles of {Nano}-{Optics}},
	copyright = {https://www.cambridge.org/core/terms},
	isbn = {978-1-108-78150-3 978-1-108-47894-6},
	url = {https://www.cambridge.org/core/product/identifier/9781108781503/type/book},
	doi = {10.1017/9781108781503},
	urldate = {2026-09-16},
	publisher = {Cambridge University Press},
	author = {Novotny, Lukas and Hecht, Bert},
	month = nov,
	year = {2025},
}

@misc{feldman_trapping_2025,
	title = {Trapping and cooling of nanodiamonds in a {Paul} trap under ultra-high vacuum: {Towards} matter-wave interferometry with massive objects},
	copyright = {Creative Commons Attribution Non Commercial No Derivatives 4.0 International},
	shorttitle = {Trapping and cooling of nanodiamonds in a {Paul} trap under ultra-high vacuum},
	url = {https://arxiv.org/abs/2508.14687},
	doi = {10.48550/ARXIV.2508.14687},
	urldate = {2026-09-17},
	publisher = {arXiv},
	author = {Feldman, Omer and Shultz, Ben Baruch and Muretova, Maria and Dobkowski, Or and Japha, Yonathan and Grosswasser, David and Folman, Ron},
	year = {2025},
	note = {Version Number: 1},
}

@article{siegel_optical_2025,
	title = {Optical levitation of arrays of microspheres},
	volume = {111},
	url = {https://link.aps.org/doi/10.1103/PhysRevA.111.033514},
	doi = {10.1103/PhysRevA.111.033514},
	number = {3},
	urldate = {2026-09-17},
	journal = {Physical Review A},
	publisher = {American Physical Society},
	author = {Siegel, Benjamin and Afek, Gadi and Lowe, Cecily and Wang, Jiaxiang and Tseng, Yu-Han and Penny, T. W. and Moore, David C.},
	month = mar,
	year = {2025},
	pages = {033514},
}

@misc{noauthor_comsol_nodate,
	address = {Stockholm, Sweden},
	title = {{COMSOL} {Multiphysics}® v. 6.1},
	url = {www.comsol.com},
	publisher = {COMSOL AB},
}

@article{hebestreit_calibration_2018,
	title = {Calibration and temperature measurement of levitated optomechanical sensors},
	volume = {89},
	issn = {0034-6748, 1089-7623},
	url = {http://arxiv.org/abs/1711.09049},
	doi = {10.1063/1.5017119},
	number = {3},
	urldate = {2026-07-24},
	journal = {Review of Scientific Instruments},
	author = {Hebestreit, Erik and Frimmer, Martin and Reimann, René and Dellago, Christoph and Ricci, Francesco and Novotny, Lukas},
	month = mar,
	year = {2018},
	note = {arXiv:1711.09049 [physics.optics]},
	pages = {033111},

}

@article{hauer_general_2013,
	title = {A general procedure for thermomechanical calibration of nano/micro-mechanical resonators},
	volume = {339},
	issn = {00034916},
	url = {http://arxiv.org/abs/1305.0557},
	doi = {10.1016/j.aop.2013.08.003},
	urldate = {2026-07-24},
	journal = {Annals of Physics},
	author = {Hauer, B. D. and Doolin, C. and Beach, K. S. D. and Davis, J. P.},
	month = dec,
	year = {2013},
	note = {arXiv:1305.0557 [cond-mat.mes-hall]},
	pages = {181--207},

}

@book{li_fundamental_2013,
	address = {New York, NY},
	series = {Springer {Theses}},
	title = {Fundamental {Tests} of {Physics} with {Optically} {Trapped} {Microspheres}},
	copyright = {https://www.springernature.com/gp/researchers/text-and-data-mining},
	isbn = {978-1-4614-6030-5 978-1-4614-6031-2},
	url = {https://link.springer.com/10.1007/978-1-4614-6031-2},
	doi = {10.1007/978-1-4614-6031-2},
	language = {english},
	urldate = {2026-07-24},
	publisher = {Springer New York},
	author = {Li, Tongcang},
	year = {2013},

}

@phdthesis{burrell_force_2025,
	title = {Force {Sensitive} {Levitated} {Optomechanics} for {Short} {Range} {Gravity}},
	copyright = {Database copyright ProQuest LLC; ProQuest does not claim copyright in the individual underlying works.},
	isbn = {979-8-3157-9839-2},
	url = {https://www.proquest.com/docview/3215569736/abstract/9289056FB159486EPQ/1},
	language = {English},
	urldate = {2026-09-17},
	school = {ProQuest Dissertations \& Theses},
	author = {Burrell, Nia},
	year = {2025},
}

@article{novotnydrop,
  title = {Sensing Static Forces with Free-Falling Nanoparticles},
  author = {Hebestreit, Erik and Frimmer, Martin and Reimann, Ren\'e and Novotny, Lukas},
  journal = {Phys. Rev. Lett.},
  volume = {121},
  issue = {6},
  pages = {063602},
  numpages = {5},
  year = {2018},
  month = {Aug},
  publisher = {American Physical Society},
  doi = {10.1103/PhysRevLett.121.063602},
  url = {https://link.aps.org/doi/10.1103/PhysRevLett.121.063602}
}

@misc{skrabulis2026nanomechanicalsensorresolvingimpulsive,
      title={Nanomechanical sensor resolving impulsive forces below its zero-point fluctuations}, 
      author={Martynas Skrabulis and Martin Colombano Sosa and Nicola Carlon Zambon and Andrei Militaru and Massimiliano Rossi and Martin Frimmer and Lukas Novotny},
      year={2026},
      eprint={2601.19392},
      archivePrefix={arXiv},
      primaryClass={quant-ph},
      url={https://arxiv.org/abs/2601.19392}, 
}

@misc{rossi2024quantumdelocalizationlevitatednanoparticle,
      title={Quantum Delocalization of a Levitated Nanoparticle}, 
      author={Massimiliano Rossi and Andrei Militaru and Nicola Carlon Zambon and Andreu Riera-Campeny and Oriol Romero-Isart and Martin Frimmer and Lukas Novotny},
      year={2024},
      eprint={2408.01264},
      archivePrefix={arXiv},
      primaryClass={quant-ph},
      url={https://arxiv.org/abs/2408.01264}, 
}

@misc{kremer2026fastquantumsqueezingnanomechanical,
      title={Fast quantum squeezing of a nanomechanical oscillator with an inverted potential}, 
      author={Oscar Schmitt Kremer and Lorenzo Dania and Lukas Novotny and Martin Frimmer},
      year={2026},
      eprint={2609.12833},
      archivePrefix={arXiv},
      primaryClass={quant-ph},
      url={https://arxiv.org/abs/2609.12833}, 
}

@article{Moore_2021,
doi = {10.1088/2058-9565/abcf8a},
url = {https://doi.org/10.1088/2058-9565/abcf8a},
year = {2021},
month = {jan},
publisher = {IOP Publishing},
volume = {6},
number = {1},
pages = {014008},
author = {Moore, David C and Geraci, Andrew A},
title = {Searching for new physics using optically levitated sensors},
journal = {Quantum Science and Technology}
}

@article{Mooredecay,  
title = {Mechanical Detection of Nuclear Decays},
  author = {Wang, Jiaxiang and Penny, T. W. and Recoaro, Juan and Siegel, Benjamin and Tseng, Yu-Han and Moore, David C.},
  journal = {Phys. Rev. Lett.},
  volume = {133},
  issue = {2},
  pages = {023602},
  numpages = {6},
  year = {2024},
  month = {Jul},
  publisher = {American Physical Society},
  doi = {10.1103/PhysRevLett.133.023602},
  url = {https://link.aps.org/doi/10.1103/PhysRevLett.133.023602}
}

@article{Vinante_2019,
   title={Testing collapse models with levitated nanoparticles: Detection challenge},
   volume={100},
   ISSN={2469-9934},
   url={http://dx.doi.org/10.1103/PhysRevA.100.012119},
   DOI={10.1103/physreva.100.012119},
   number={1},
   journal={Physical Review A},
   publisher={American Physical Society (APS)},
   author={Vinante, A. and Pontin, A. and Rashid, M. and Toroš, M. and Barker, P. F. and Ulbricht, H.},
   year={2019},
   month=jul }

@article{Hamaide2026,
  title = {Searching for ultralight dark matter with MOLeQuTE: A massive optically levitated quantum tabletop experiment},
  author = {Hamaide, Louis and Banks, Hannah and Barker, Peter and Geraci, Andrew A.},
  journal = {Phys. Rev. Res.},
  pages = {},
  year = {2026},
  month = {Jul},
  publisher = {American Physical Society},
  doi = {10.1103/2t8j-3wzj},
  url = {https://link.aps.org/doi/10.1103/2t8j-3wzj}
}

@article{Arvanitaki:2013,
  title = {Detecting High-Frequency Gravitational Waves with Optically Levitated Sensors},
  author = {Arvanitaki, Asimina and Geraci, Andrew A.},
  journal = {Phys. Rev. Lett.},
  volume = {110},
  issue = {7},
  pages = {071105},
  numpages = {5},
  year = {2013},
  month = {Feb},
  publisher = {American Physical Society},
  doi = {10.1103/PhysRevLett.110.071105},
  url = {https://link.aps.org/doi/10.1103/PhysRevLett.110.071105}
}

@article{aggarwal2022searching,
  title={Searching for new physics with a levitated-sensor-based gravitational-wave detector},
  author={Aggarwal, Nancy and Winstone, George P and Teo, Mae and Baryakhtar, Masha and Larson, Shane L and Kalogera, Vicky and Geraci, Andrew A},
  journal={Physical Review Letters},
  volume={128},
  number={11},
  pages={111101},
  year={2022},
  publisher={APS}
}

@article{ORI11_GM,
   author = {Romero-Isart, Oriol},
   title = {Quantum superposition of massive objects and collapse models},
   journal = {Phys. Rev. A},
   volume = {84},   
   pages = {052121},
   DOI = {10.1103/PhysRevA.84.052121},
   url = {https://link.aps.org/doi/10.1103/PhysRevA.84.052121},
   year = {2011},
   type = {Journal Article}
}

@Article{Bateman2014,
author={Bateman, James
and Nimmrichter, Stefan
and Hornberger, Klaus
and Ulbricht, Hendrik},
title={Near-field interferometry of a free-falling nanoparticle from a point-like source},
journal={Nature Communications},
year={2014},
month={Sep},
day={02},
volume={5},
number={1},
pages={4788},
issn={2041-1723},
doi={10.1038/ncomms5788},
url={https://doi.org/10.1038/ncomms5788}
}

@article{Goldman2015,
  title = {Sensing short range forces with a nanosphere matter-wave interferometer},
  author = {Geraci, Andrew and Goldman, Hart},
  journal = {Phys. Rev. D},
  volume = {92},
  issue = {6},
  pages = {062002},
  numpages = {7},
  year = {2015},
  month = {Sep},
  publisher = {American Physical Society},
  doi = {10.1103/PhysRevD.92.062002},
  url = {https://link.aps.org/doi/10.1103/PhysRevD.92.062002}
}

@misc{deplano2026stationaryentanglementlevitatedoscillator,
      title={Stationary entanglement of a levitated oscillator with an optical field}, 
      author={Q. Deplano and A. Pontin and F. Marino and F. Marin},
      year={2026},
      eprint={2602.03456},
      archivePrefix={arXiv},
      primaryClass={quant-ph},
      url={https://arxiv.org/abs/2602.03456}, 
}

@article{Mancini2003,
  title = {Scheme for Teleportation of Quantum States onto a Mechanical Resonator},
  author = {Mancini, Stefano and Vitali, David and Tombesi, Paolo},
  journal = {Phys. Rev. Lett.},
  volume = {90},
  issue = {13},
  pages = {137901},
  numpages = {4},
  year = {2003},
  month = {Apr},
  publisher = {American Physical Society},
  doi = {10.1103/PhysRevLett.90.137901},
  url = {https://link.aps.org/doi/10.1103/PhysRevLett.90.137901}
}

@article{RMPiontrapnoise,
  title = {Ion-trap measurements of electric-field noise near surfaces},
  author = {Brownnutt, M. and Kumph, M. and Rabl, P. and Blatt, R.},
  journal = {Rev. Mod. Phys.},
  volume = {87},
  issue = {4},
  pages = {1419--1482},
  numpages = {64},
  year = {2015},
  month = {Dec},
  publisher = {American Physical Society},
  doi = {10.1103/RevModPhys.87.1419},
  url = {https://link.aps.org/doi/10.1103/RevModPhys.87.1419}
}

@Article{Abbasov2023,
author={Abbasov, T.
and Zibrov, S.
and Sherstov, I.},
title={Surface-Electrode Ion Trap Development},
journal={JETP Letters},
year={2023},
month={Aug},
day={01},
volume={118},
number={3},
pages={215-219},
issn={1090-6487},
doi={10.1134/S0021364023602063},
url={https://doi.org/10.1134/S0021364023602063}
}

@article{patch,
  title = {Measuring the effect of electrostatic patch potentials in Casimir force experiments},
  author = {Garrett, Joseph L. and Kim, Jongbum and Munday, Jeremy N.},
  journal = {Phys. Rev. Research},
  volume = {2},
  issue = {2},
  pages = {023355},
  numpages = {5},
  year = {2020},
  month = {Jun},
  publisher = {American Physical Society},
  doi = {10.1103/PhysRevResearch.2.023355},
  OPTurl = {https://link.aps.org/doi/10.1103/PhysRevResearch.2.023355}
}

@article{tao_single-crystal_2014,
	title = {Single-crystal diamond nanomechanical resonators with quality factors exceeding one million},
	volume = {5},
	issn = {2041-1723},
	url = {https://www.nature.com/articles/ncomms4638},
	doi = {10.1038/ncomms4638},
	language = {english},
	number = {1},
	urldate = {2026-09-21},
	journal = {Nature Communications},
	author = {Tao, Y. and Boss, J. M. and Moores, B. A. and Degen, C. L.},
	month = apr,
	year = {2014},
	pages = {3638},
}

@article{stowe_attonewton_1997,
	title = {Attonewton force detection using ultrathin silicon cantilevers},
	volume = {71},
	issn = {0003-6951, 1077-3118},
	url = {https://pubs.aip.org/apl/article/71/2/288/68067/Attonewton-force-detection-using-ultrathin-silicon},
	doi = {10.1063/1.119522},
	language = {english},
	number = {2},
	urldate = {2026-09-21},
	journal = {Applied Physics Letters},
	author = {Stowe, T. D. and Yasumura, K. and Kenny, T. W. and Botkin, D. and Wago, K. and Rugar, D.},
	month = jul,
	year = {1997},
	pages = {288--290},
}

@article{rugar_improved_1989,
	title = {Improved fiber‐optic interferometer for atomic force microscopy},
	volume = {55},
	issn = {0003-6951},
	url = {https://doi.org/10.1063/1.101987},
	doi = {10.1063/1.101987},
	number = {25},
	urldate = {2026-09-21},
	journal = {Applied Physics Letters},
	author = {Rugar, D. and Mamin, H. J. and Guethner, P.},
	month = dec,
	year = {1989},
	pages = {2588--2590},
}

@article{talbot_thermophoresis_1980,
	title = {Thermophoresis of particles in a heated boundary layer},
	volume = {101},
	copyright = {https://www.cambridge.org/core/terms},
	issn = {0022-1120, 1469-7645},
	url = {https://www.cambridge.org/core/product/identifier/S0022112080001905/type/journal_article},
	doi = {10.1017/S0022112080001905},
	language = {english},
	number = {4},
	urldate = {2026-09-21},
	journal = {Journal of Fluid Mechanics},
	author = {Talbot, L. and Cheng, R. K. and Schefer, R. W. and Willis, D. R.},
	month = dec,
	year = {1980},
	pages = {737--758},
}

@article{fuchs_aerosols,

title = {The mechanics of aerosols. By N. A. Fuchs. Translated by R. E. Daisley and Marina Fuchs; Edited by C. N. Davies. London (Pergamon Press), 1964. Pp. xiv, 408; 82 Figures; 40 Tables. £6},
journal = {Quarterly Journal of the Royal Meteorological Society},
volume = {91},
number = {388},
pages = {249-249},
author = {Fuchs, N. A.},
doi = {https://doi.org/10.1002/qj.49709138822},
url = {https://rmets.onlinelibrary.wiley.com/doi/abs/10.1002/qj.49709138822},
eprint = {},
year = {1965}
}

@article{rider_electrically_2019,
	title = {Electrically driven, optically levitated microscopic rotors},
	volume = {99},
	url = {https://link.aps.org/doi/10.1103/PhysRevA.99.041802},
	doi = {10.1103/PhysRevA.99.041802},
	number = {4},
	urldate = {2026-09-21},
	journal = {Physical Review A},
	publisher = {American Physical Society},
	author = {Rider, Alexander D. and Blakemore, Charles P. and Kawasaki, Akio and Priel, Nadav and Roy, Sandip and Gratta, Giorgio},
	month = apr,
	year = {2019},
	pages = {041802},
}

@article{blakemore_absolute_2020,
	title = {Absolute pressure and gas species identification with an optically levitated rotor},
	volume = {38},
	issn = {2166-2746},
	url = {https://doi.org/10.1116/1.5139638},
	doi = {10.1116/1.5139638},
	number = {2},
	urldate = {2026-09-21},
	journal = {Journal of Vacuum Science \& Technology B},
	author = {Blakemore, Charles P. and Martin, Denzal and Fieguth, Alexander and Kawasaki, Akio and Priel, Nadav and Rider, Alexander D. and Gratta, Giorgio},
	month = jan,
	year = {2020},
	pages = {024201},
}

@article{rider_search_2016,
	title = {Search for {Screened} {Interactions} {Associated} with {Dark} {Energy} below the \$100{\textbackslash}text\{ \}{\textbackslash}ensuremath\{{\textbackslash}mu\}{\textbackslash}mathrm\{m\}\$ {Length} {Scale}},
	volume = {117},
	url = {https://link.aps.org/doi/10.1103/PhysRevLett.117.101101},
	doi = {10.1103/PhysRevLett.117.101101},
	number = {10},
	urldate = {2026-09-21},
	journal = {Physical Review Letters},
	publisher = {American Physical Society},
	author = {Rider, Alexander D. and Moore, David C. and Blakemore, Charles P. and Louis, Maxime and Lu, Marie and Gratta, Giorgio},
	month = aug,
	year = {2016},
	pages = {101101},
}

@article{crookes_xv_1874,
	title = {{XV}. {On} attraction and repulsion resulting from radiation},
	copyright = {https://royalsociety.org/journals/ethics-policies/data-sharing-mining/},
	issn = {0261-0523, 2053-9223},
	url = {https://royalsocietypublishing.org/rstl/article/doi/10.1098/rstl.1874.0015/119029/XV-On-attraction-and-repulsion-resulting-from},
	doi = {10.1098/rstl.1874.0015},
	language = {english},
	number = {164},
	urldate = {2026-09-21},
	journal = {Philosophical Transactions of the Royal Society of London},
	author = {Crookes, William},
	month = dec,
	year = {1874},
	pages = {501--527},
}

@article{tebben_optimaldetection,
  title = {Optimal position detection of a dipolar scatterer in a focused field},
  author = {Tebbenjohanns, Felix and Frimmer, Martin and Novotny, Lukas},
  journal = {Phys. Rev. A},
  volume = {100},
  issue = {4},
  pages = {043821},
  numpages = {10},
  year = {2019},
  month = {Oct},
  publisher = {American Physical Society},
  doi = {10.1103/PhysRevA.100.043821},
  url = {https://link.aps.org/doi/10.1103/PhysRevA.100.043821}
}

@article{mamin_sub-attonewton_2001,
	title = {Sub-attonewton force detection at millikelvin temperatures},
	volume = {79},
	issn = {0003-6951},
	url = {https://doi.org/10.1063/1.1418256},
	doi = {10.1063/1.1418256},
	number = {20},
	journal = {Applied Physics Letters},
	author = {Mamin, H. J. and Rugar, D.},
	month = nov,
	year = {2001},
	pages = {3358--3360},
}

\end{document}